\documentclass[referee]{aa}
\usepackage{graphicx}

\def\kpc{{\ h^{-1} \ \rm kpc}}
\def\msun{{\ h^{-1}  \ \rm M_{\odot}}}

\begin{document}

   \title{Galaxy pairs in cosmological simulations: effects of interactions
on colours and chemical abundances}

   \author{M. Josefa Perez, 
          \inst{1}\inst{2}\inst{3}
           Patricia B. Tissera,
          \inst{1}\inst{3}
           Cecilia Scannapieco
          \inst{1}\inst{3}
	   Diego G. Lambas
                  \inst{1}\inst{4}
          \and
	  Maria E. De Rossi
          \inst{1}\inst{3}
                    }

   \offprints{M. Josefa Perez}

   \institute{Consejo Nacional de Investigaciones Cient\'{\i}ficas y T\'ecnicas.
         \and 
              Facultad de Ciencias Astron\'omicas y Geof\'{\i}sicas, La Plata\\
              \email{jperez@fcaglp.unlp.edu.ar}
         \and 
              Instituto de Astronom\'{\i}a y F\'{\i}sica del Espacio, Argentina.\\
              \email{patricia@iafe.uba.ar}
         \and 
              Observatorio Astron\'omico de la Universidad Nacional de C\'ordoba, Argentina.\\
             }


   \abstract{
We perform an statistical analysis of galaxies in pairs in a $\Lambda$-CDM scenario by using
the chemical {\small GADGET-2} of Scannapieco et al. (2005) in order to study the effects of galaxy interactions
on colours and metallicities. We   find that galaxy-galaxy interactions can
produce a bimodal colour distribution with  galaxies with significant recent star formation activity
contributing mainly to  blue colours.  In the simulations,  the colours and the fractions
of recently formed stars of  galaxies in pairs depend on environment more strongly than those of galaxies
without a close companion, suggesting that interactions play an important role in
galaxy evolution. 
If the metallicity of the stellar populations is used as the chemical indicator, we find that the simulated galaxies  determine 
  luminosity-metallicity and  stellar mass-metallicity relations which do not depend on  the presence of a close companion. 
However, in the case of the luminosity-metallicity relation, at a given level of enrichment,
 we detect a systematic displacement of the
relation to brighter magnitudes for active star forming systems.
Regardless of relative distance and current level of star formation activity, galaxies in pairs have stellar
populations with higher level of enrichment than galaxies without a close companion. In the case of
the gas component, this is no longer valid for galaxies in pairs with passive star formation which 
only show an excess of metals for very close pair members, consequence of an important recent past star formation activity.
In agreement with observations, the signature of gas inflows driven by interactions can be also detected
in the lower mean O/H abundances measured in the central regions of galaxies in pairs.
Our results suggest that interactions play a significant role in the determination of 
 colour and chemical properties of galaxies in hierarchical clustering scenarios, although SN energy feedback is needed to achieve
a full agreement.

   \keywords{cosmology: theory - galaxies: formation -
galaxies: evolution - galaxies: interactions.}
   }
            
   \authorrunning{M. Josefa Perez et al}
   
   \titlerunning{Colours and metallicities in galaxy pairs }
   
   \maketitle
%

\section{Introduction}

Colours and chemical abundances are two important parameters that characterize galaxies.
The combination of stellar populations with different ages and metallicities produces colour distributions 
with particular features and correlations such as the luminosity-metallicity relation. 
Each
galaxy has an unique evolutionary history which sets  the pace of the transformation of gas into stars
by establishing the collapse time, the accretion rates and the impact of
 interactions and mergers. On its turn, these processes regulate the metal production  which can 
modify the cooling rates, affecting the subsequent gas accretion and
star formation activity. 
Other important factors that could contribute to determine the astrophysical, dynamical and chemical
properties of galaxies are environmental effects such as the action of 
tidal stripping and ram-pressure as well as  
interactions and mergers.
Disentangling the effects of each of these  processes is extremely difficult. One sucessful approach
is to carry out statistical analysis of large galaxy samples with the aim at underpinning their main characteristics
and their dependence on the astrophysical properties of galaxies.

Large galaxy surveys such as the 2dF Galaxy Redshift Survey (2dFGRS) and the  
Sloan Digital Sky Survey (SDSS)  have provided new insights
into the properties of galaxies in the local universe. It is accepted 
that galaxies  are grouped in two general types: blue or red
(Baldry et al. 2004 and Kauffmann et al.2004), disc-dominated or spheroid-dominated,
active or passive star forming. 
In particular, recent detailed analysis of the colour distributions of galaxies
by Baldry et al.(2004) and Balogh et al. (2004, hereafter B04) have shown that these distributions can
be well described by the combination of two Gaussians over an important
range of magnitudes and densities. While the  
Gaussians features are found to depend on luminosity so that  brighter  
systems have  redder colours, the peak locations seem not to change
with local density. It is only the fraction of galaxies in the blue
and in the red distributions which are found to vary with local density, with
high density regions
 mostly populated by red galaxies. 
 According to B04, the invariability they
found in the two Gaussian distributions with local density suggests that the
responsible process of transforming blue into red galaxies has to be very efficient 
to overcome the effects of environment. 
Recently, Driver et al. (2006) also analysed the colour distribution of galaxies in pairs in 
the Millenium Galaxy Catalog. They confirmed that spheroid-type galaxies tend to populate the red
distribution while disc-dominated ones determine the blue distribution. Hence, these results suggest
that the physical mechanism
responsible of the colour properties could be also linked to a change in galaxy morphology.
One possible mechanism is galaxy-galaxy interactions.

As several observational works have shown that the  proximity to a companion can be statistically related to
an enhancement of the star formation activity (Barton et al. 2000; Lambas et al. 2003; Nikolic et al. 2003),
 regardless of environment (Alonso et al. 2005, 2006). The results found in these works agree with
the theoretical understanding that interactions can drive strong torques which may trigger gas inflows
and violent star formation (e.g. Martinet 1995).
Kewley, Geller \& Barton (2005) also studied galaxies in pairs but focusing on the effects of
interactions on the metallicity properties of these galaxies. At a given luminosity, they  found
 a trend for a lower  
 enrichment in the central regions of galaxies with a close neighbour than in those of field galaxies.
 The relation between  luminosity and metallicity (LMR) is  well-known in the local  universe being 
 followed from dwarf galaxies to
elliptical galaxies. Recent works from the SDSS have confirmed with outstanding statistical level the slope and zero point
of this correlation at $z=0$ (Tremonti et al. 2004 and references therein).
When luminosity is substitued by stellar mass a more fundamental relation appears. The mass-metallicity relation
(MMR)
was first determined by Lequeux et al. (1979) and then confirmed by Tremonti et al. (2004).
New observations at high redsfhit suggest an evolution for both the LMR and MMR in both zero point and slope
  (Kobulnicky \& Kewley 2004;
Liang et al. 2004; Savaglio et al. 2005; Erb et al. 2006). An statistical analysis of the metallicity properties of galaxies in pairs is still
missing (Alonso et al., in preparation).

In the current cosmological paradigm, galaxies formed by the aggregation of substructure, implying that mergers and
interactions are common events in galaxy formation. Many important works have been advocated to the study of
the effects of interactions and mergers by using numerical simulations as first discussed by
Toomre \& Toomre (1972). Numerical simulations have been improved
over time in order to  include more complex physical processes such as hydrodynamics, radiative cooling and star formation
(e.g. Thomas \& Couchmann 1992; Navarro \& White 1994; Tissera, Lambas \& Abadi 1997; Springel \& Hernquist 2002). 
Galaxy-galaxy interactions have been studied by several authors who  found  that
tidal torques which developed during  interactions could be strong and efficient enough to produce non-axisymmetrical
instabilities in disc systems (e.g. Athanassoula \& Sellwood 1986; Binney \& Tremaine 1987;
Barnes \& Hernquist 1991, 1992; Christodoulou, Shlosman \& Tohline 1995). As a consequence, gas inflows
can be triggered feeding important new star formation activity. 
Tissera et al. (2002) showed that, in hierarchical clustering scenarios, these mechanisms acted with different
degree of efficiency along the evolutionary paths of galaxies, contributing to the formation of compact
stellar bulges. These spheroidal components help to stabilize the systems.
Later on, Scannapieco \& Tissera (2003) showed that interactions and mergers could drive a morphological loop. 
 According to these works, depending on
the history of evolution of galaxies, systems may react in a different way during a merger event 
depending on their  internal properties.

 P\'erez et al. (2006, hereafter Paper I) investigated the effects of galaxy-galaxy interactions
in the concordance cosmological framework. Numerical simulations
were perfomed
by using the  chemical GADGET-2 version of
Scannapieco et al. (2005). 
 In Paper I, we studied the effects of interactions on the star formation (SF) activity of galaxies analyzing its
dependence on orbital parameters and local density. 
We found that simulated galaxies showed an enhancement
of the star formation activity with  proximity to a companion. This trend was proved to be
in agreement with observational findings of Barton et al. (2000) and Lambas et al. (2003).
In Paper I, it was also shown that proximity together with the
properties of the potential well of the systems played an important role in the triggering of
SF by interactions. We also detected that  interactions accelerate the evolutionary process producing a larger
fraction of passively, more stable systems in pairs, in agreement with recent observational results of
Alonso et al. (2004, 2006). 

In this paper, we continue the analysis of galaxy pairs in the $\Lambda$-CDM scenario discussed in
Paper I extending our investigations to colours and metallicities.
Recently, chemical evolution has been included in hydrodyamical simulations (e.g. Mosconi et al. 2001;
Lia et al. 2001; Scannapieco et al. 2005) allowing the consistent description of the chemical
enrichment of the baryonic matter as galaxies are assembled. Chemical properties can store fossil
records of the interacting systems which can help us to confront models with observations.
In this work, we   focus on trying to unveil the effects of interactions on the colour
distributions and metallicity properties of galaxies in pairs in the Local Universe.

This paper is organized as follows. In Section 2 we describe briefly the main characteristics of the numerical simulations and
the criteria used to select the galaxy pairs. In Section 3, we discuss the role
 of interactions on the colour distribution of galaxies and compare our results with the observational ones.
The chemical properties of galaxy pairs are analysed in Section 4. Section 5 summarizes our main findings.

\section{Simulated galaxy pair catalogs}

For our analysis, we use the galaxy pair (GP) catalogs defined in Paper I. 
Here we only summarize the main characteristics of the simulations and of the catalogs  referring
the interested reader to Scannapieco et al. (2005) and Perez et al. (2005) for more details on
the chemical code and pair selection, respectively.

The GP catalogs  were constructed from a 10 Mpc $h^{-1}$ cubic volume of   a $\Lambda$-CDM Universe
($\Lambda=0.7$, $\Omega=0.3$, $H_0=100 h$ ${\rm km \ s^{-1} Mpc^{-1}}$ with $ h=0.7$) 
run with  the chemical  code of Scannapieco 
et al. (2005) developed in  {\small GADGET-2} (Springel \& Hernquist 2003). 
A total of $2\times 80^3$ particles were used, yielding a 
dark matter  mass of $M_{\rm DM}= 1.4 \times 10^{8} 
{\rm M_{\odot}} h^{-1}$ and an initial mass for gas particles of 
$M_{\rm gas}= 2.2\times 10^{7} {\rm M_{\odot}} h^{-1}$. 
The chemical {\small GADGET-2}  describes   
the enrichment of the interstellar medium by Supernovae Type II (SNII)  and Type Ia (SNIa).
In this simulation, a Salpeter Initial Mass Function has been adopted with 
 0.1 and 40$ {\rm M_{\odot}}$ cut-offs. The theoretical yields of Woosley \& Weaver (2005) and
Thielemann et al. (1993) have been used for SNII and SNIa chemical production, respectively.
Note that these simulations are consistent with an instantaneous thermalization of the SN energy which
has  no
impact on the dynamics and the star formation history of the structure (e.g. Katz 1992).

Galactic systems were identified by using a fine-tunning friends-of-friends algorithm within virialized 
structures to select all baryonic bounded  clumps (see Paper I for more details). A minimum stellar mass of $8\times 10^8 \msun $ for the clumps was imposed to minimize resolution problems. With this low mass cut-off, a total
of 364 galactic systems were identified.
 
In Paper I, galaxy pairs were selected 
 based only on
a proximity criterium for two main reasons. First, for the sake of comparison with 
observational galaxy-pair analyses
that used proximity
criterium to select pairs  (e.g. Barton et al. 2000; Lambas et al. 2001;  Alonso et al. 2006). Secondly,
because we are interested in the effects of galaxy-galaxy interactions even if they
 do not end being a merger event.

We used the tridimensional  simulated galaxy pair catalog (3D-GP)  constructed  in Paper I  where
a  
  distance  criterium  $r < 100 \kpc$ was estimated but requesting the galaxy members
to exhibit enhanced star formation activity with respect to the average of the simulated galaxy sample, following
the observational procedure first explained by 
Lambas et al. (2001).
The 3D-GP catalog is made up of  88 galaxy pairs. 

We also analysed the projected galaxy pair (2D-GP) catalog built  up by projecting the tridimensional 
total galaxy distribution according to three random directions.
 The 2D-GP catalog comprises 677 galaxies in pairs
with 
projected separation  $r_{\rm p}<100 \kpc$  and radial velocity  $\Delta cz <350$ km s$^{-1}$. 
The comparison between results from the 3D-GP and 2D-GP catalogs allows the determination of the effects
of spurious pairs formed only by projection effects. A detailed discussion of this point can be
found in Paper I.

We also define the control samples for both catalogs made up
of  galaxies without a close companion within the corresponding relative distance and velocity thresholds.
The comparison of the properties of galaxies in pairs with those of galaxies without a close
companion selected from the same simulation 
allows us to unveil  the effects of interactions  as  shown by Lambas et al. (2001).

The GP catalogs provide information on 
 the stellar birthrate parameter
$b$ (defined as the present level of SF activity of a galaxy
normalized to its mean past SF rate), the total stellar mass $M_*$, the fraction of  stars formed in the last 0.5 Gyr
$F^*$ and  the mean stellar age $\tau$
 of all  simulated galaxies within two optical radii. The optical radius is defined as the one
that encloses 83 per cent of the baryonic matter of the system.
We classify active and passive SF galaxies as those which exhibit higher or lower star formation activity
in comparison with the average $b$ parameter of the control sample ($b =0.8$).

For each simulated galaxy,  we also estimate the mean oxygen abundance (O/H) and the absolute magnitudes 
 by considering stars (or gas) within two optical
radii.
Absolute magnitudes in different wavelenghts
 have been estimated by combining the star formation rate histories and the
metal content of each simulated stellar population and convolving this information with
the population synthesis models of Bruzual \& Charlot (2003) as explained in more detail 
by De Rossi et al. (in preparation).

 In order to characterize the environment of galaxies, we use the local projected density $\Sigma$ defined
as $\Sigma=6/(\pi d^{2})$, where {\it d} is  the projected distance
to the 6$^{th}$ nearest neighbour brighter than M$_{\rm r} = -20.5 $  and with 
$\Delta cz<1000$ km ${\rm s^{-1}}$ (these limits have been taken from B04 and have been already used in Paper I).

\section{Colour distribution in tridimensional galaxy pairs}
 
\begin{figure}
\centering
\includegraphics[width=8cm,height=6.5cm]{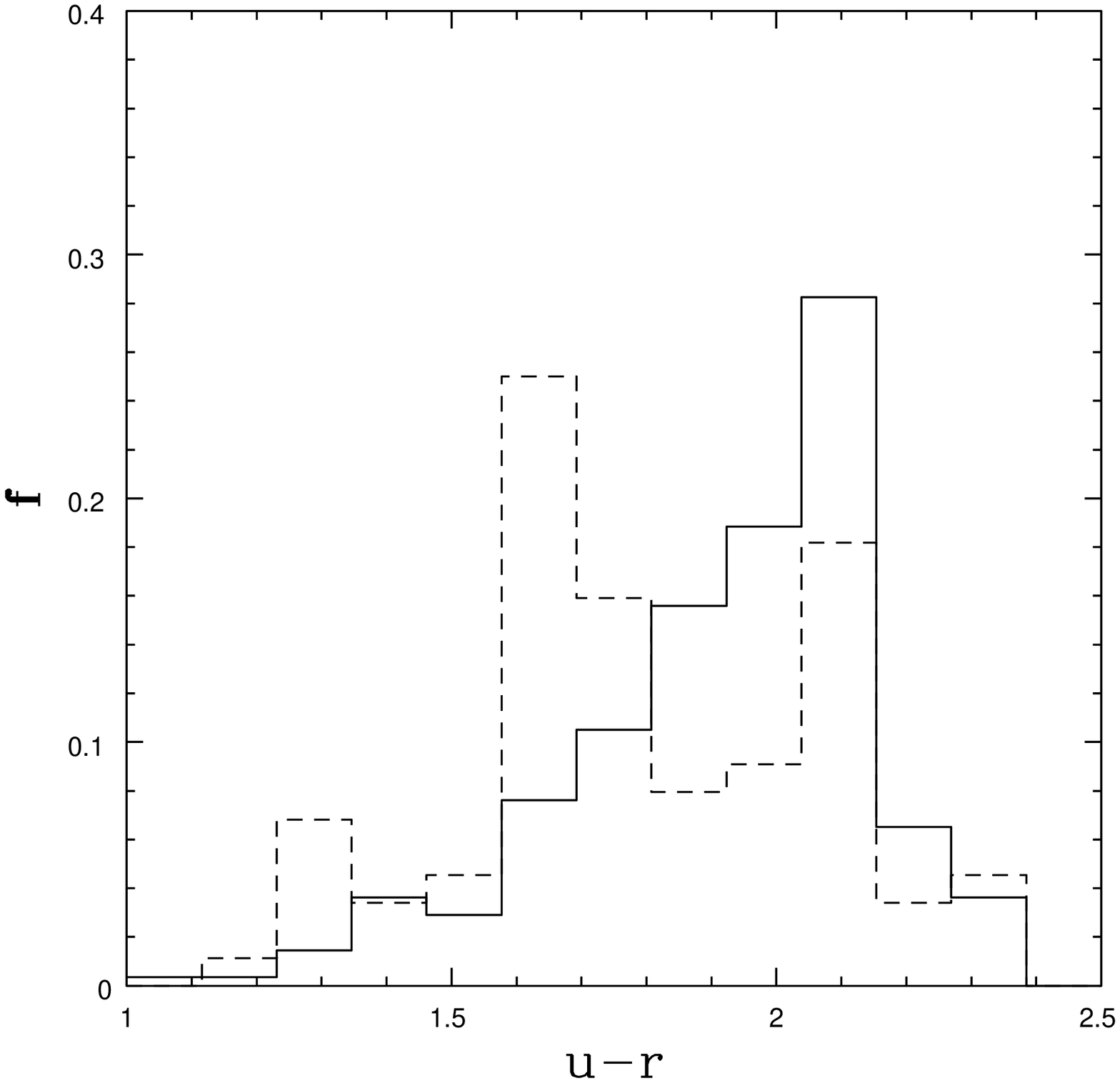}
\includegraphics[width=8cm,height=6.5cm]{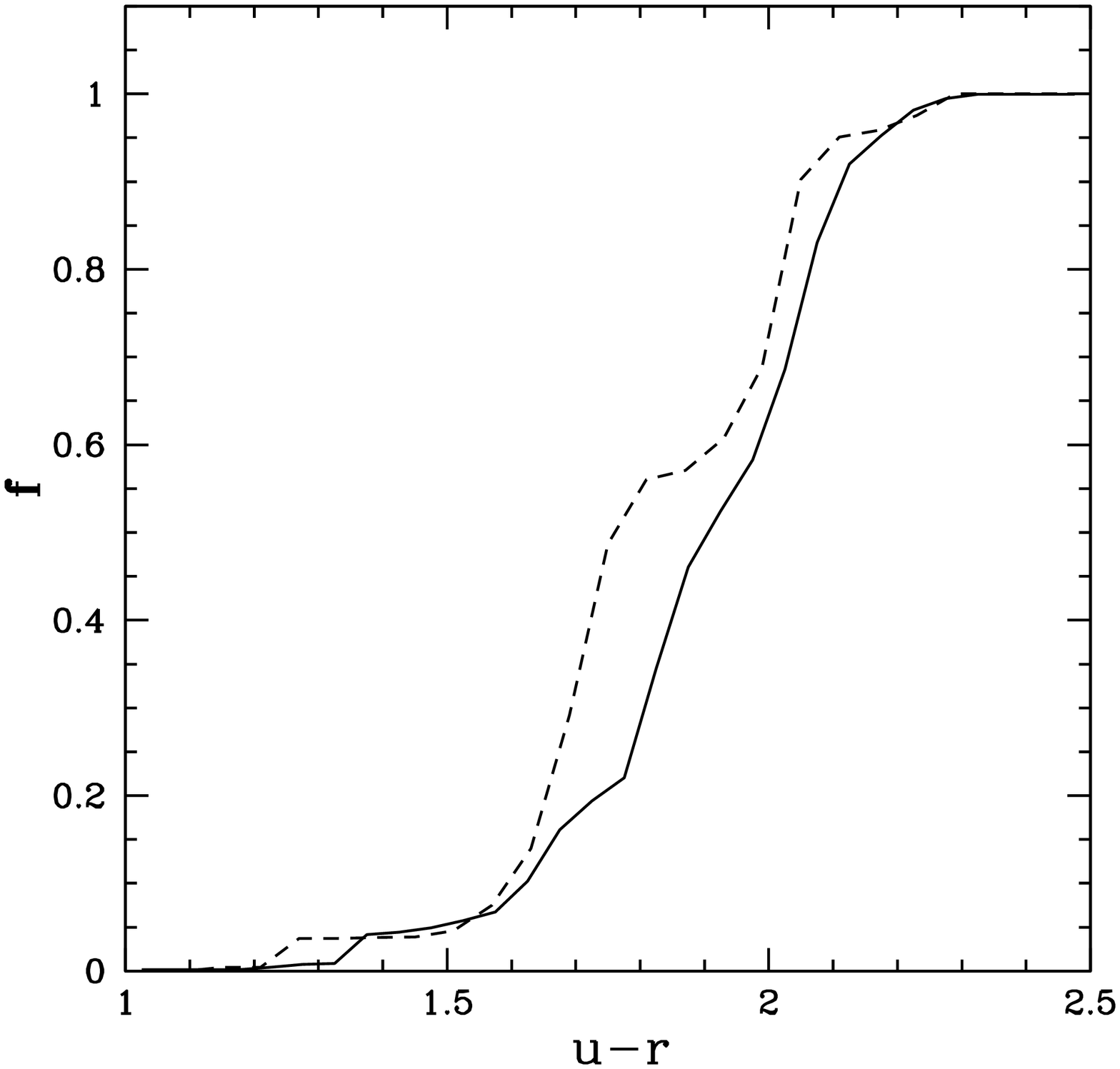}
\caption{Upper panel: Colour distribution for galaxies in tridimensional pairs (dashed line) and in the 
corresponding control 
sample (solid line). Lower panel:
Cumulative fraction of the total stellar mass contained in
these galaxy pairs (dashed line) and in the control sample (solid line) as a function
of $u - r$ colour. }
\label{colour-pares}
\end{figure}

In this Section, we analyse the $u-r$ colour distribution of the simulated galactic systems
for our 3D-GP sample. In order to infer the  effect of interactions on the colour
distribution, we also analyse the corresponding one for  galaxies without a near
companion (control sample).
In Fig.~\ref{colour-pares} we can appreciate the colour distributions of  galaxies in pairs 
and galaxies without a close companion displayed as histograms  (upper panel) and as stellar mass partition
functions (lower panel).
From the results of 
B04,  we adopt the value $u - r =1.8$ to roughly define  the limit between the red and blue distributions.
We found that the mean colours of the blue and red peaks for simulated pairs (dashed line)
 are at $<u -r > \approx 1.6$ and 
$<u -r > \approx 2.1$, with $58$ per cent of galaxies in the blue distribution. 
The control sample (solid line)  has similar values:   $<u -r > \approx 1.6$ and  $<u -r > \approx 2.0$, but with $26$ 
per cent of galaxies in the blue distribution.
As we can see from this figure, simulated galaxies in pairs  have a clear bimodal colour distribution. 
However,
the red peaks of the simulated galaxies in pairs and in the control sample 
are displaced toward blue colours by $\approx 0.20$ dex with respect to observations. For similar
environment, B04 reported the blue mean at  $<u -r > \approx 1.5$ and the red mean at  $<u -r > \approx 2.3$.

In  Fig.~\ref{colour-pares} (lower panel),  we show  
the cumulative fraction of stellar mass in galaxies in pairs (dashed line) and in the control sample (solid line)
 as a function of colour.  We  appreciate that 
while $\approx 60 $ per cent of the stellar mass in galaxies in pairs contribute to the blue colours ($u - r < 1.8$),
only $\approx 30$ per cent  of isolated galaxies  do so. 
According to the analysis of Kauffmann et al. (2004),
blue galaxies contribute to
$\approx 45 $ per cent  of the stellar mass, averaging over  all environments.   
 Hence for the control galaxies, we found a red excess ($ \approx 15$ per cent) 
  which could be related to 
 the  high efficiency in the transformation of gas into stars  in our simulations.
Because of this,  the cold and dense gas  is quickly converted in stars,  producing the reddening of
 systems.
The treatment of SN energy feedback could help to regulate the SF activity preventing the
artifially efficient SF activity  leading to a colour distribution
in full agreement with observations. However, the description of SN feedback in SPH codes
is still a controversial issue and only recently a novel scheme has been presented by
Scannapieco et al. (2006) which seems to be well-suited for galaxy formation in a cosmological context. 

In order to get insight into the origin of the bimodal colour
 distribution obtained for  the simulated pair sample,
 following Paper I  we   split the galaxy pairs
 into merging ($r<30 \kpc$) and  interacting  
($30 \kpc <r<100 \kpc$) pairs. Note that the merging subsample has been named considering that their members
are very close pairs and that they are   physically interacting even if they are not gravitationally
 bounded systems.

\begin{figure}
\centering
\includegraphics[width=8cm,height=6.5cm]{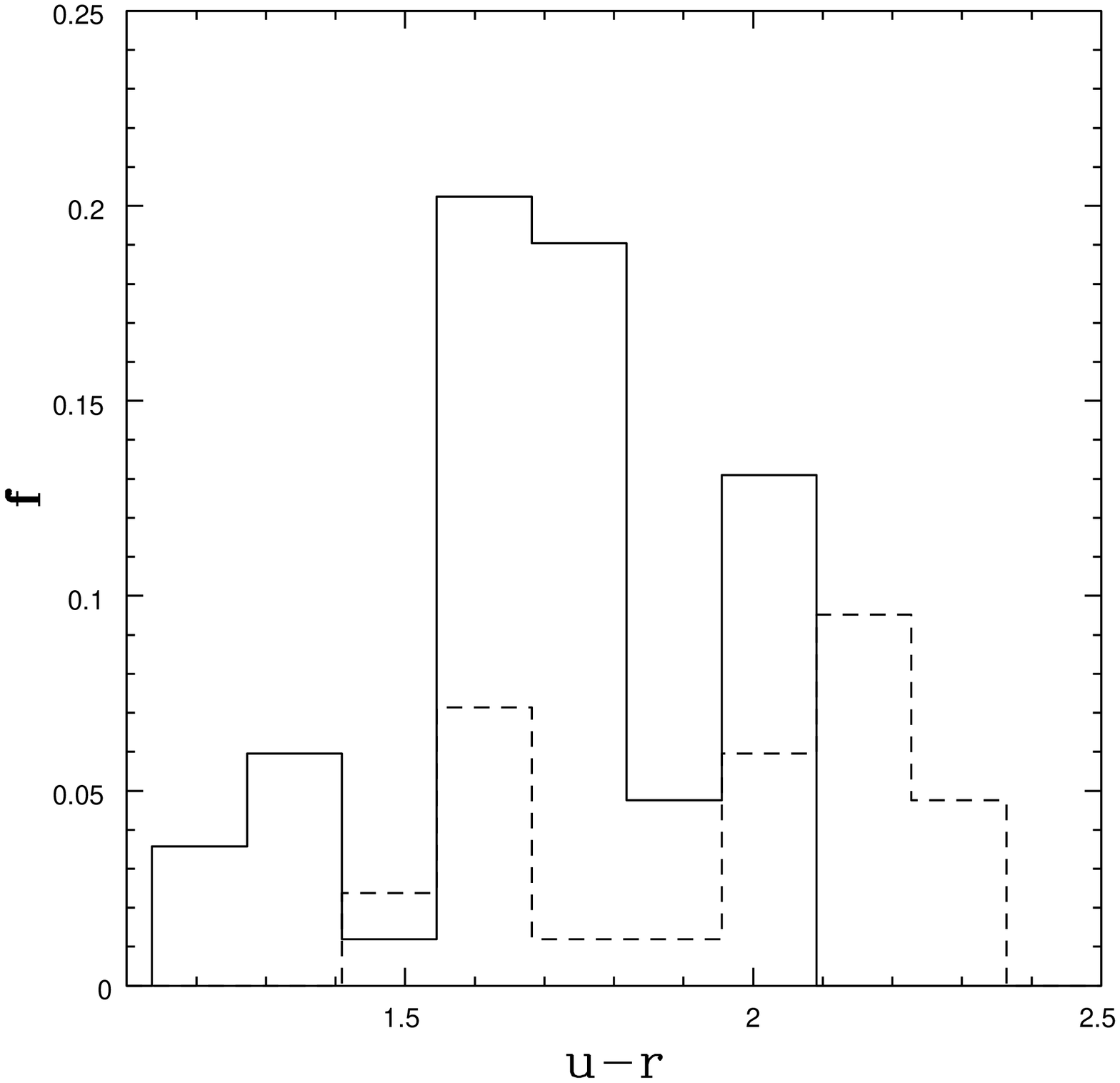}\\
\includegraphics[width=8cm,height=6.5cm]{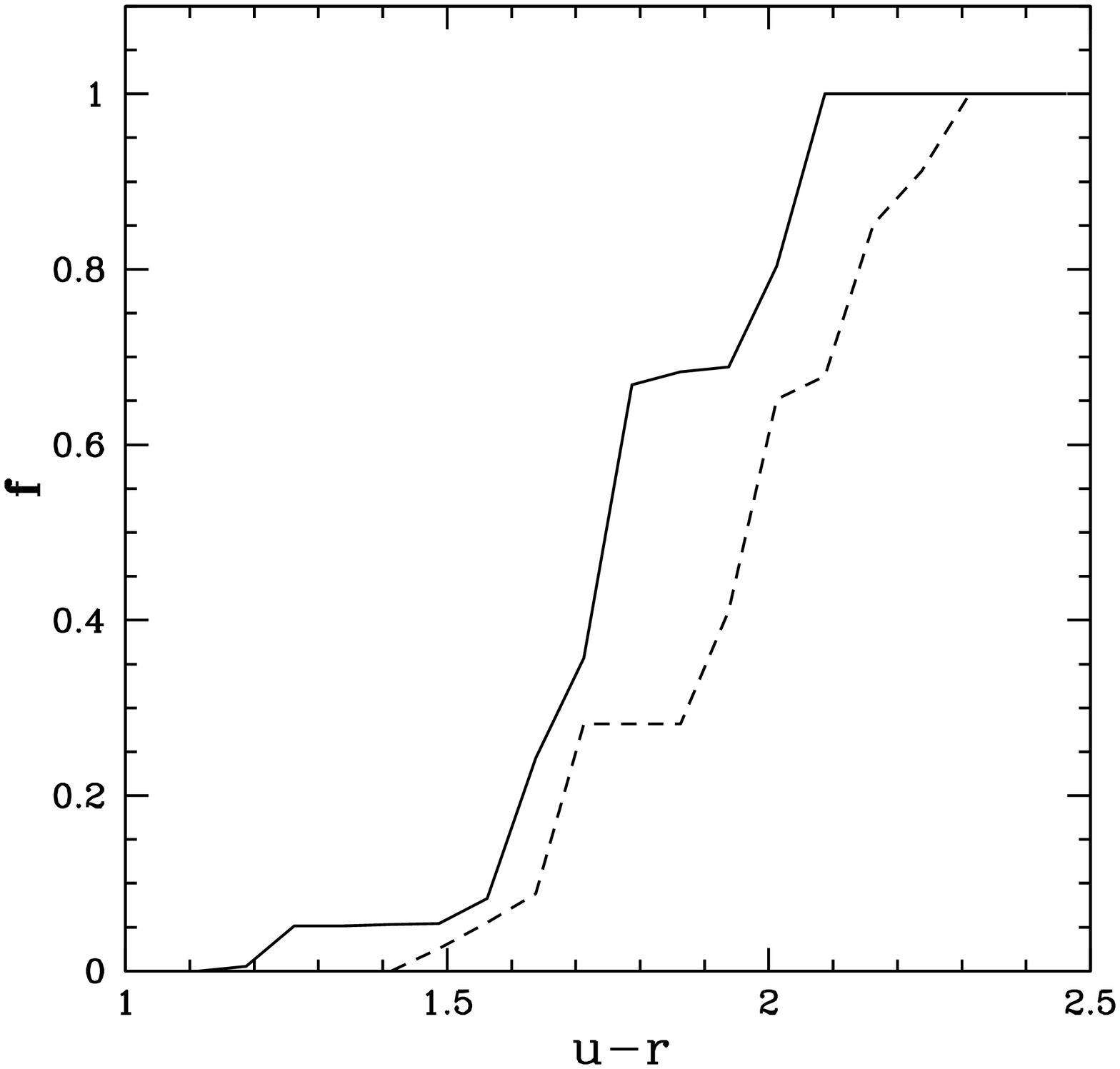}
\caption{Upper panel: Colour distribution for galaxies in 
tridimensional pairs separating in merging 
( $r < 30 \kpc$, solid line) and
interacting ($30 \kpc < r < 100 \kpc$ dashed line) systems. Lower panel:
Cumulative fraction of the total stellar mass contained in
merging galaxies (solid line) and in interacting ones (dashed line) as a function
of $u - r$ colour. }
\label{close-interacting}
\end{figure}

\begin{figure}
\centering
\includegraphics[width=8cm,height=6.5cm]{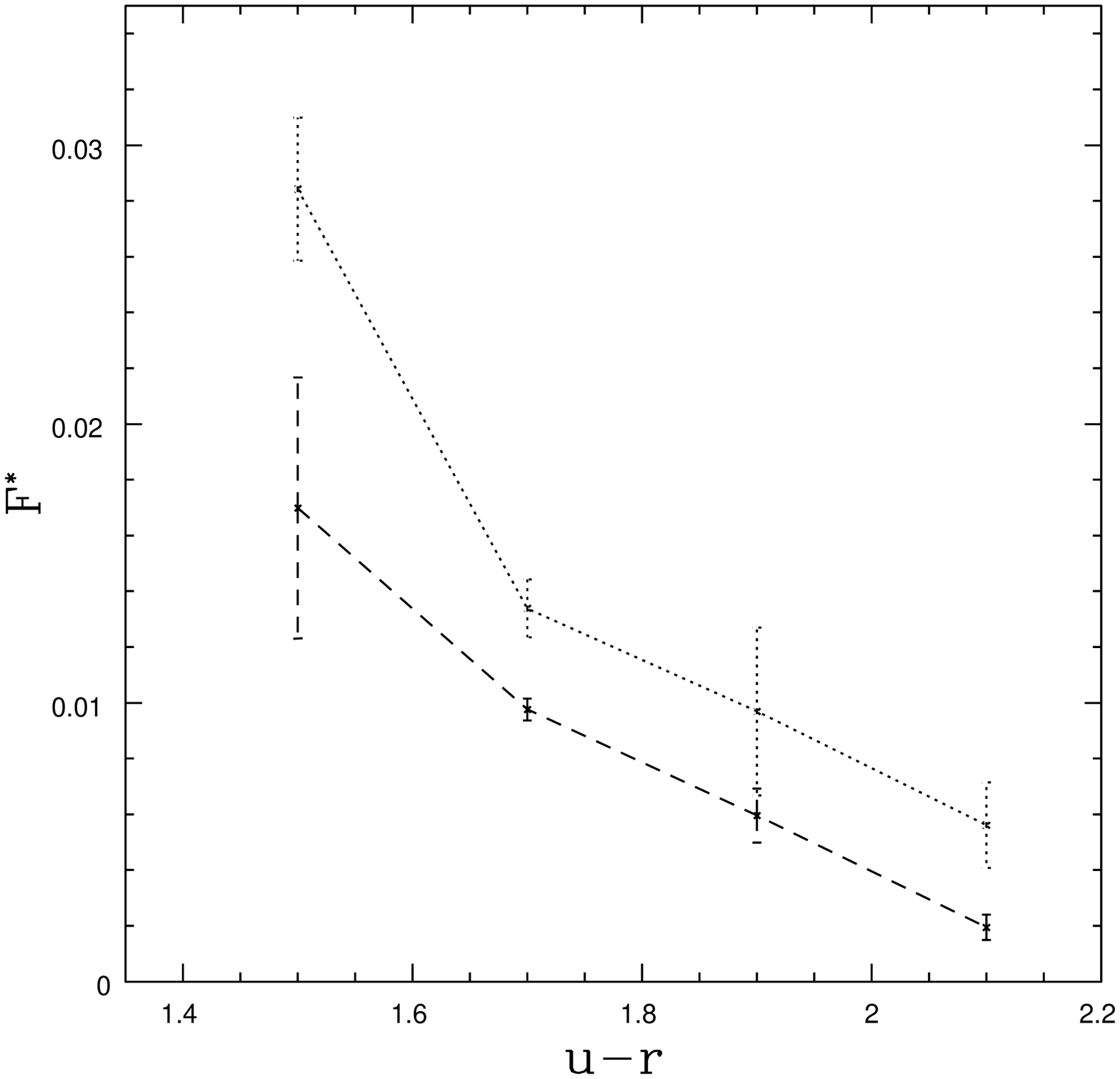}
\caption{Fraction of stars formed in the last $0.5$ Gyr as a function of
$u - r$ colour for active   (dotted line) and  passive  (dashed lines)
star forming   galaxies in  tridimensional pairs. Error bars have been estimated by
applying the  bootstrap resampling technique. }
\label{fraccion-star}
\end{figure}

In Fig.~\ref{close-interacting} we can appreciate the stellar mass partition function (lower panel) 
and colour histograms (upper panel) for  both subsamples. 
From the colour histograms (upper panel), we estimate that 70 per cent of galaxies in merging pairs
has colours  $u-r < 1.8$, while only  33 per cent of interacting pairs has colours in this range.
This trend is also shown by the cumulative fraction of the total stellar mass as a function of colour (lower
panel).  
Merging pairs have $\approx 65 $ per cent of their stellar mass
in systems with $u-r < 1.8$. In the case of interacting pairs, only $\approx 30 $ per cent of the stellar mass
is associated to blue systems.
From Paper I we know that 40 per cent of galaxies in merging pairs have enhanced SF activity with
respect to galaxies without a close companion. For interacting pairs, 25 per cent of galaxies
exhibit such level of SF activity. It was also proved in Paper I that for passive SF galaxies in 
merging systems, the fraction $F^*$ of newly born stars increased with proximity to a companion, 
evidence of the higher SF level recently experienced by these systems. 
This analysis indicated that interactions are effective at triggering SF activity. 
These recently
new born stars also contribute to make bluer colours as we can see from Fig.~\ref{fraccion-star}
where we
  show the fraction of recently formed stars $F^*$ in currently active  (dotted lines) and
passive (dashed lines) SF galaxy pairs  as a function of  the 
 $u - r$ colour. 
As we can 
see, the contribution to the blue colours  comes from both active and passive SF galaxies,
 as long as they have experienced important star formation in the last 0.5 Gyr.
A similar relation is found if pairs are divided into merging and interacting pairs, although
the former have always  larger  $F^*$ fractions.

These findings show that  galaxy-galaxy interactions are able to produce  bimodal colour distributions
in hierarchical clustering scenarios. Galaxies that contribute to the blue colours are those that are currently
forming stars at significant high rates and those that have recently experienced strong SF activity.

\section{ Properties of  galaxy  pairs in projection}

\begin{figure}
\centering
\includegraphics[width=8cm,height=6.5cm]{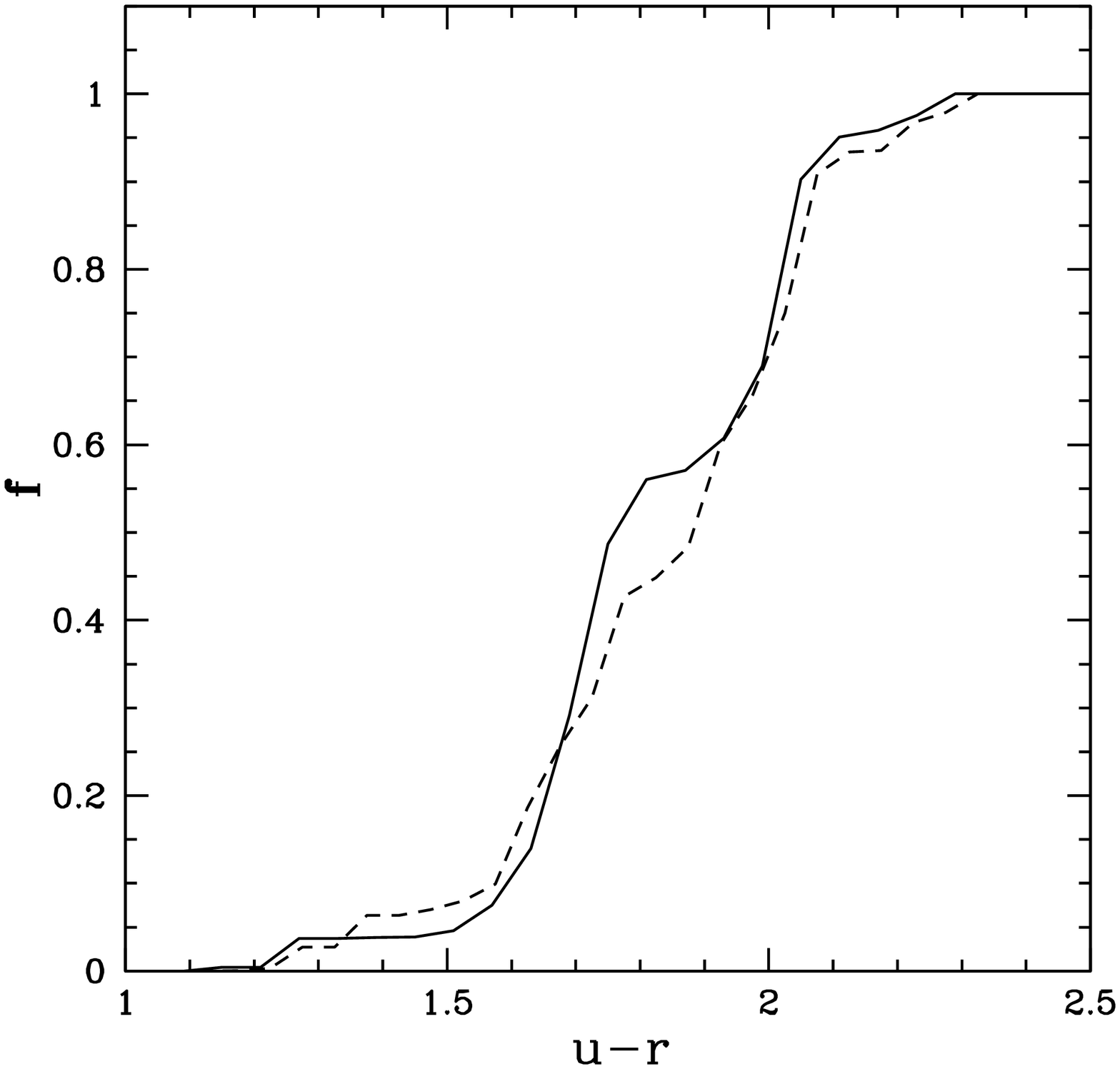}\\
\includegraphics[width=8cm,height=6.5cm]{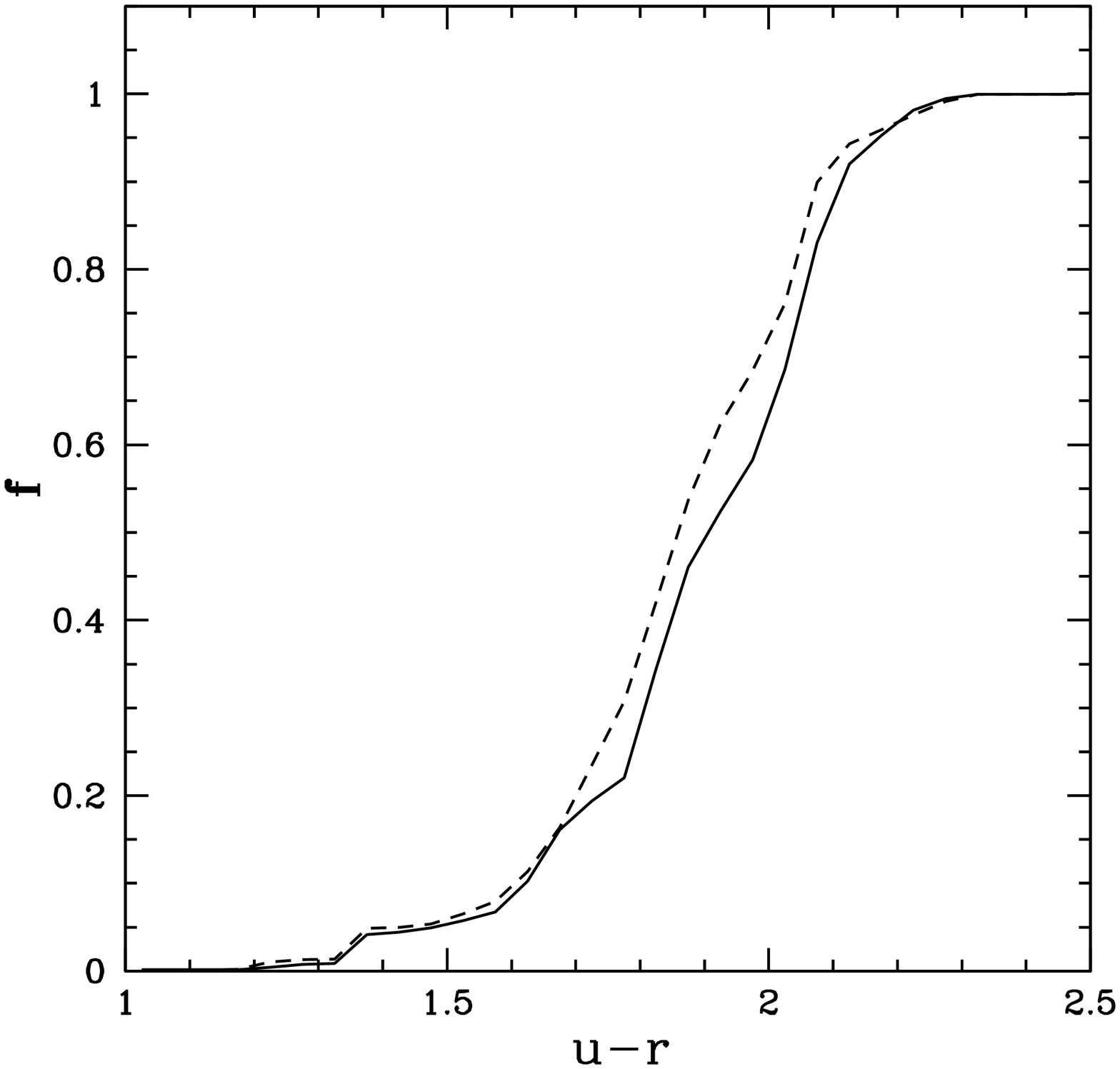}
\caption{Cumulative fraction of the total stellar mass contained in galaxies as a function of $u-r$ colour,
 in pairs (upper panel) and in the control samples (lower panel)
constructed from the tridimensional (solid lines) and projected (dashed lines) total simulated galaxy catalogs.}
\label{spurious}
\end{figure}

In order to directly compare the  astrophysical properties of simulated galaxies in pairs  
  with observations,  we use the 2D-GP catalog.
In  Paper I,  we estimated
a $30$ per cent of  interlopers  in  pairs with  $r_{\rm p}<100 \kpc$ and $\Delta cz< 350 $ km s$^{-1}$.
This percentage reduces  to  $19$ per cent for  merging systems
($r_{p}<25 \kpc $ and $\Delta cz<100 $ km s$^{-1}$).
These findings are in agreement with those found by Nikolic et al. (2004) who
 studied the projection effects on  galaxy pairs from the SDSS spectroscopic and photometric data,
estimating 
a $20  $ per cent  of  spurious systems in  pairs with  $r_{\rm p}<100 \kpc $.
We found that these fractions of interlopers do not significantly affect the results that could be derived
from the 2D catalogs.
For example, Fig.~\ref{spurious} shows the 2D (dashed lines) and 3D (solid lines) cumulative
fraction of stellar mass in 
 galaxies in pairs 
(upper panel) and in the control samples (lower panel) as function of $u-r$ colour.
As we can appreciate, the differences are less than $\approx 10$ per cent, suggesting that 
 effects of projection do not change strongly
the colour distribution of either sample.
Therefore, hereafter we continue the analysis of the properties of the 2D-GP catalog
because of its high statistical number.

\subsection{Dependence of colours and metallicities of galaxies in pairs  on local density.}

In this section, we analyse the dependence of colours and metallicities of galaxies in pairs  on environment. The sample 
has been  divided into a
low and  a high density subsamples by adopting the density threshold  
 log $\Sigma =-0.8$, and the  magnitude limit of $M_r =-18$ (B04) used in Paper I.

\begin{figure}
\centering
\includegraphics[width=8cm,height=6.5cm]{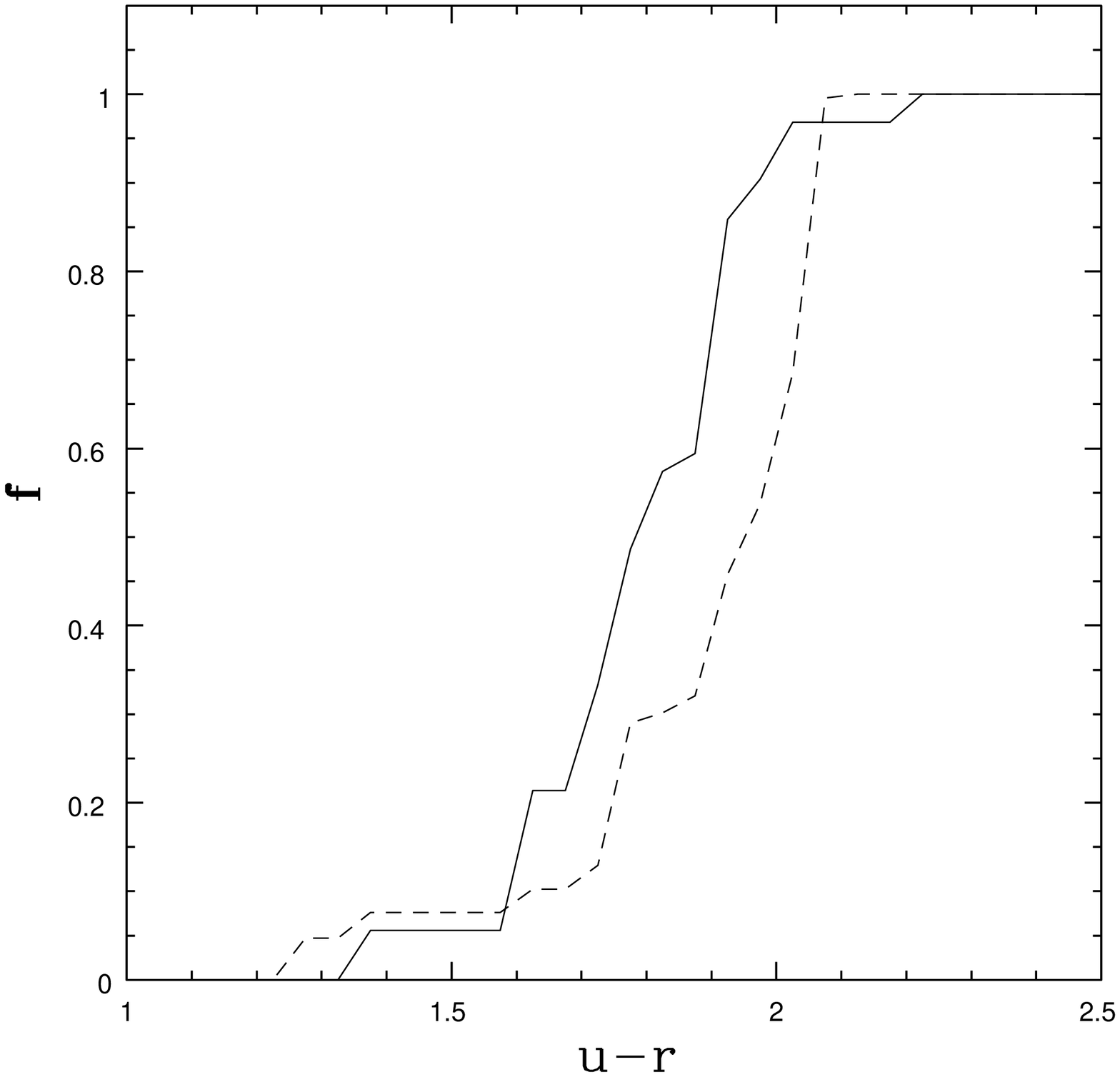}
\includegraphics[width=8cm,height=6.5cm]{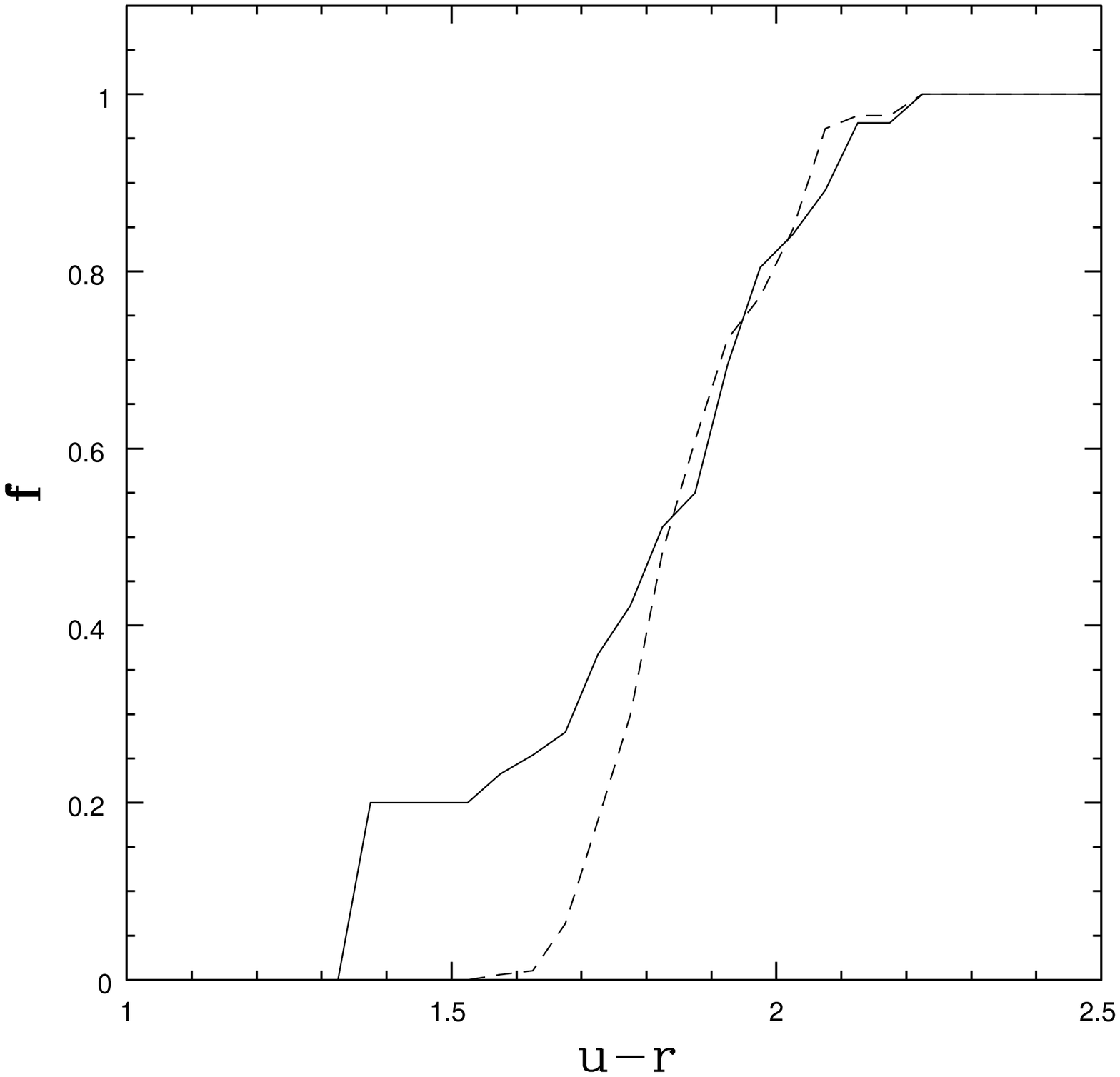}
\caption{Cumulative fraction of the total 
stellar mass contained in galaxies as a function of u-r colour.
Projected galaxies in pairs (upper panel) and in the control samples (lower panel) have been
separated in low (solid lines) and high (dashed lines) 
density regions.}
\label{acuenvi}
\end{figure}

We estimated the cumulative fraction of the total stellar mass contained in galaxies with and without a close companion
as a function of colour in the two defined environments.
As  seen from Fig. ~\ref{acuenvi},
 we  found  the expected trend for   a larger fraction of stars contributing to build up red galaxies in pairs and
in the control sample in high density regions. This trend is in agreement with the results found by 
Kauffmann et al. (2004) for the SDSS catalog. 
 
From Fig. ~\ref{acuenvi}, we can also see that the fraction of stellar mass contributing to  
red galaxies increases more strongly with local density  if the systems are in pairs.
In high density regions, a  fraction of $\approx 0.70$  of
the  stellar mass in pairs is found in systems with colours redder than $u -r = 1.8$  while this
fraction decreases to 0.55  in low density regions.
For galaxies without a close companion the fractions  of stellar mass contributing to
colours redder than  $u -r = 1.8$ are very similar in both regions.
Nevertheless, the effects of environment for the control sample  can be seen in the  absence of stellar
mass in galaxies with $u - r  < 1.6$ in  high density regions.

\begin{figure*}
\centering
\includegraphics[width=7cm,height=5.5cm]{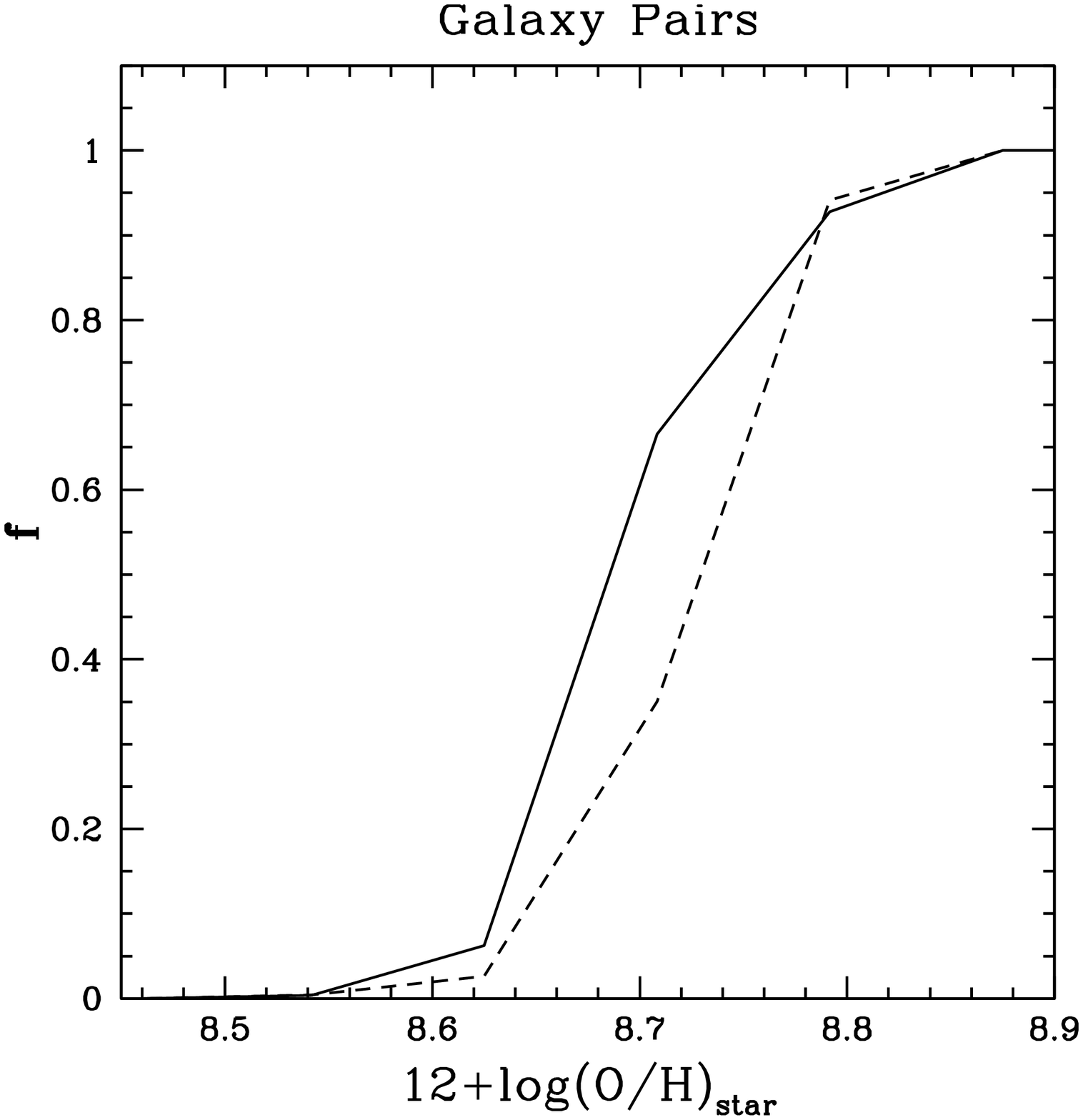}
\includegraphics[width=7cm,height=5.5cm]{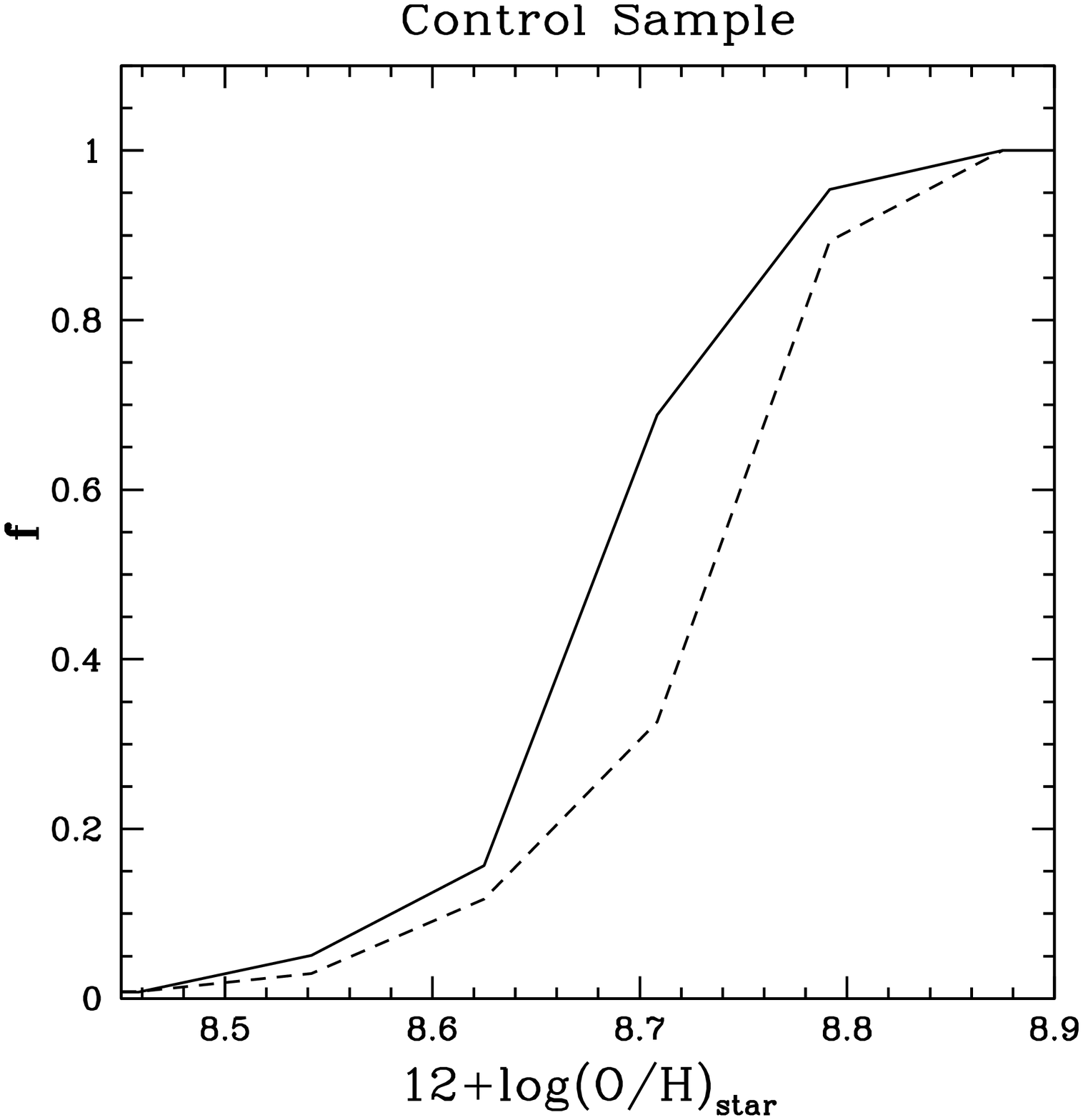}\\
\includegraphics[width=7cm,height=5.5cm]{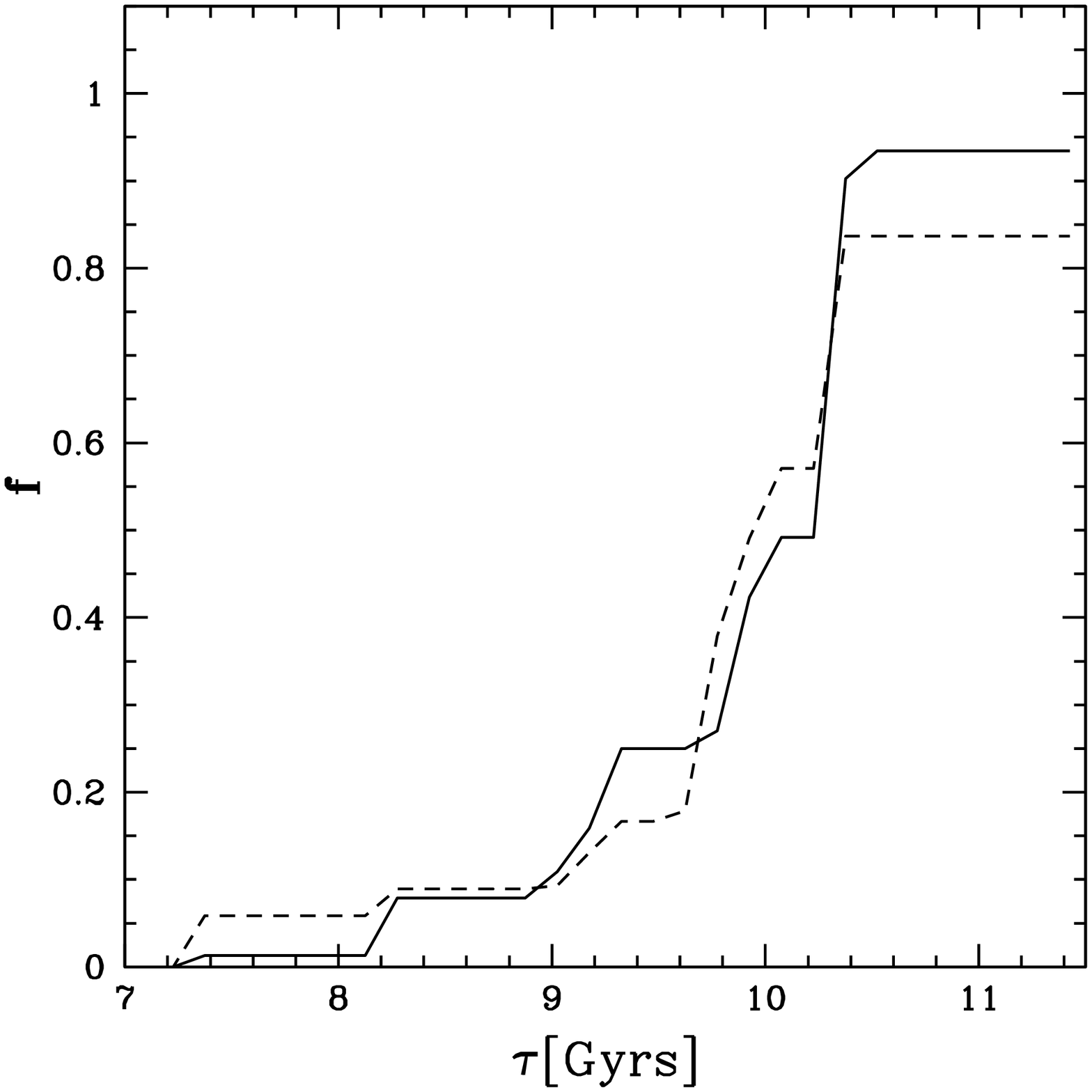}
\includegraphics[width=7cm,height=5.5cm]{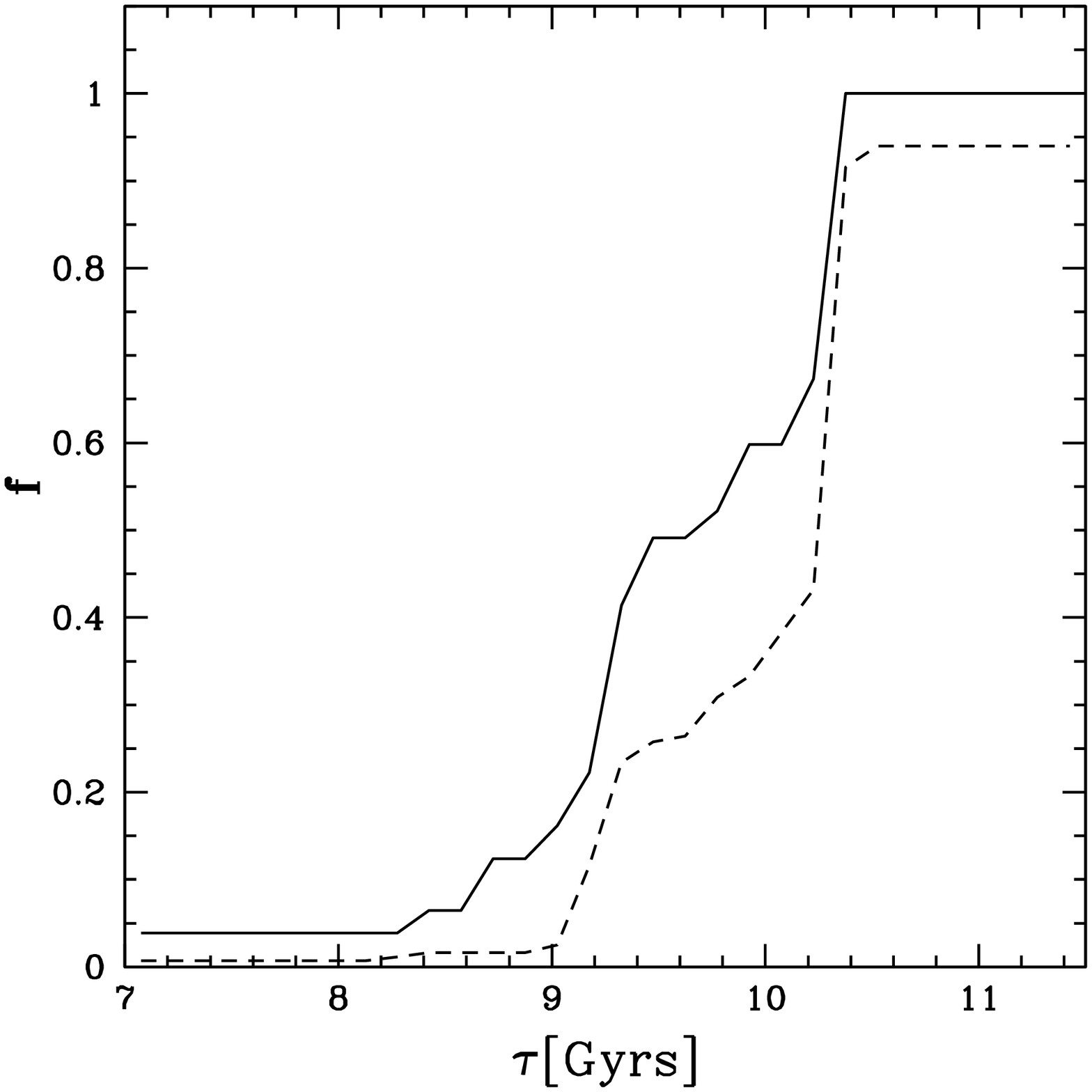}\\
\includegraphics[width=7cm,height=5.5cm]{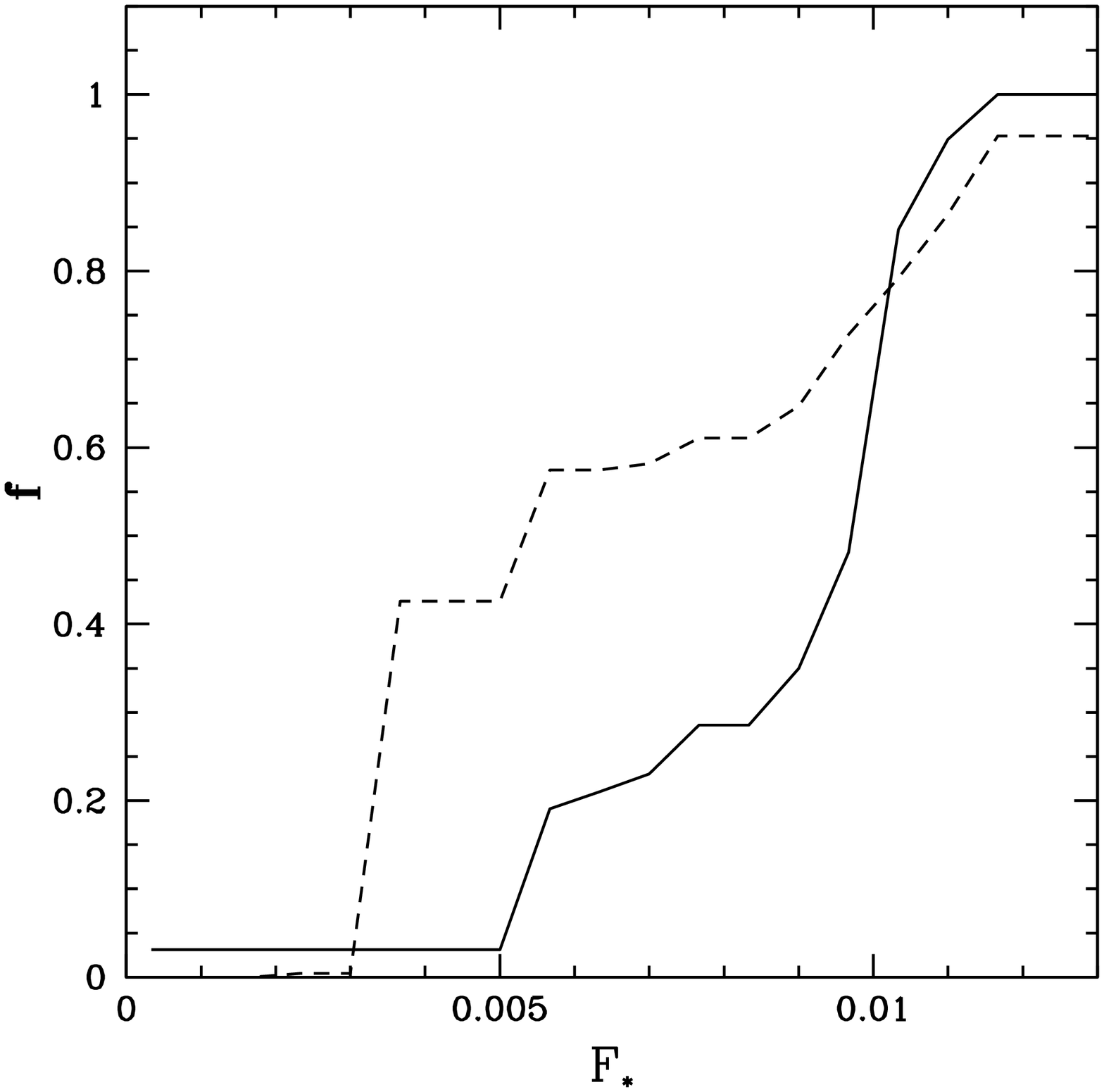}
\includegraphics[width=7cm,height=5.5cm]{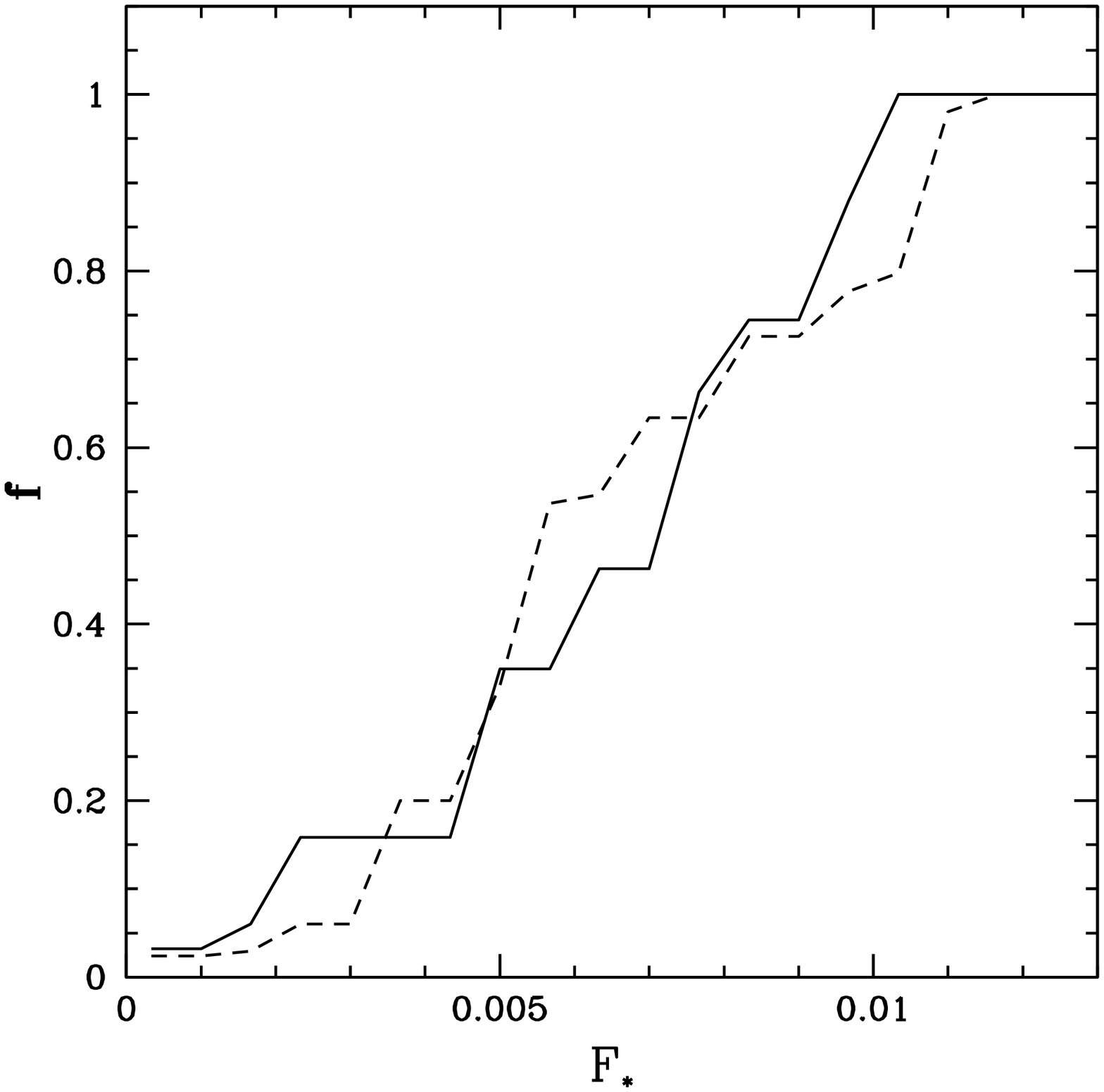}
\caption{Cumulative fraction of the total 
stellar mass contained in galaxies as a function of the  mean metallicity of the stars ($12 +$ log (O/H)),
the  mean stellar age ($\tau$) and the fraction ($F^{*}$) of stars 
formed in the last 0.5 Gyr (from top to bottom),
 for projected galaxies in pairs (left column) and in the control sample (right column). Both samples have been
separated in low (solid lines) and high (dashed lines) 
density regions.}
\label{acum}
\end{figure*}

In Fig.~\ref{acum} we
estimate the cumulative fraction of stellar mass for galaxies in pairs as a function of the mean O/H abundance
 of the
stars (a), the mean stellar age (b) and the fraction $F^*$ of recenty new born stars (c), for high and low density environments.
Similar estimations were carried out for galaxies in the control sample
(right panels).

In the case of the mean stellar age, 
galaxies in the control sample  have  a younger  underlying stellar population in low
density regions, with  $\approx 60$ per cent of the stellar mass younger than $10 $ Gyr.
In high density regions, this percentage goes down to $35$ per cent.
Galaxies in pairs have  underlying stellar populations with similar mean ages in both
environments with   $\approx 50$ per cent 
of the stellar mass having  a mean age of less than $10$ Gyr  (consistently
with the trend of galaxies in the control sample in low environments).

The mean metallicity of the global stellar population 
shows  a similar trend for galaxies in pairs and in the control samples.
Galaxies  without a close companion contribute with more stellar mass to the low
metallicity end (by approximately a factor of two) than galaxies in pairs.
Finally, the fraction of recently new born stars ($F^*$) unveils the effects of interactions and its dependence on 
environment very clearly. Galaxies in pairs in low density regions have  larger fractions of
recently formed stars with respect to their counterparts in higher density environments
and to galaxies without a close companion.

Overall, we can appreciate that,  in low density regions, galaxies in pairs
 tend to be  less metal-enriched and to have a significant
higher amount of recently born stars with respect to pairs in high density environments.

\subsection{Chemical properties of galaxies in pairs}

The chemical properties of the stellar populations (SP) 
and the interstellar medium (ISM) 
can store information on the evolution of  galaxies. 
In particular,  mergers and interactions could imprint metallicity features that
might provide  valuable information about their evolution.

\begin{figure}
\centering
\includegraphics[width=8cm,height=6.5cm]{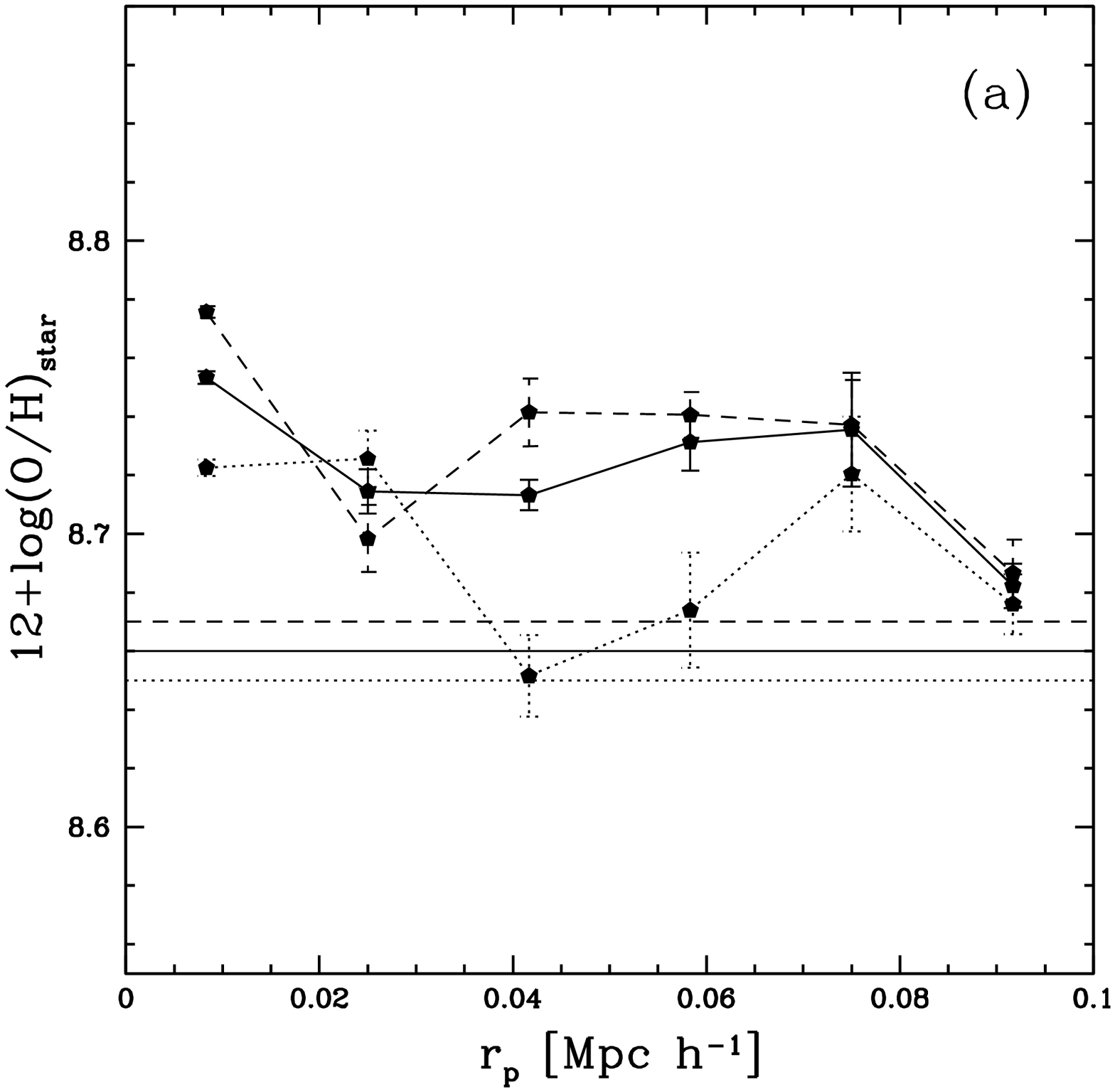}
\includegraphics[width=8cm,height=6.5cm]{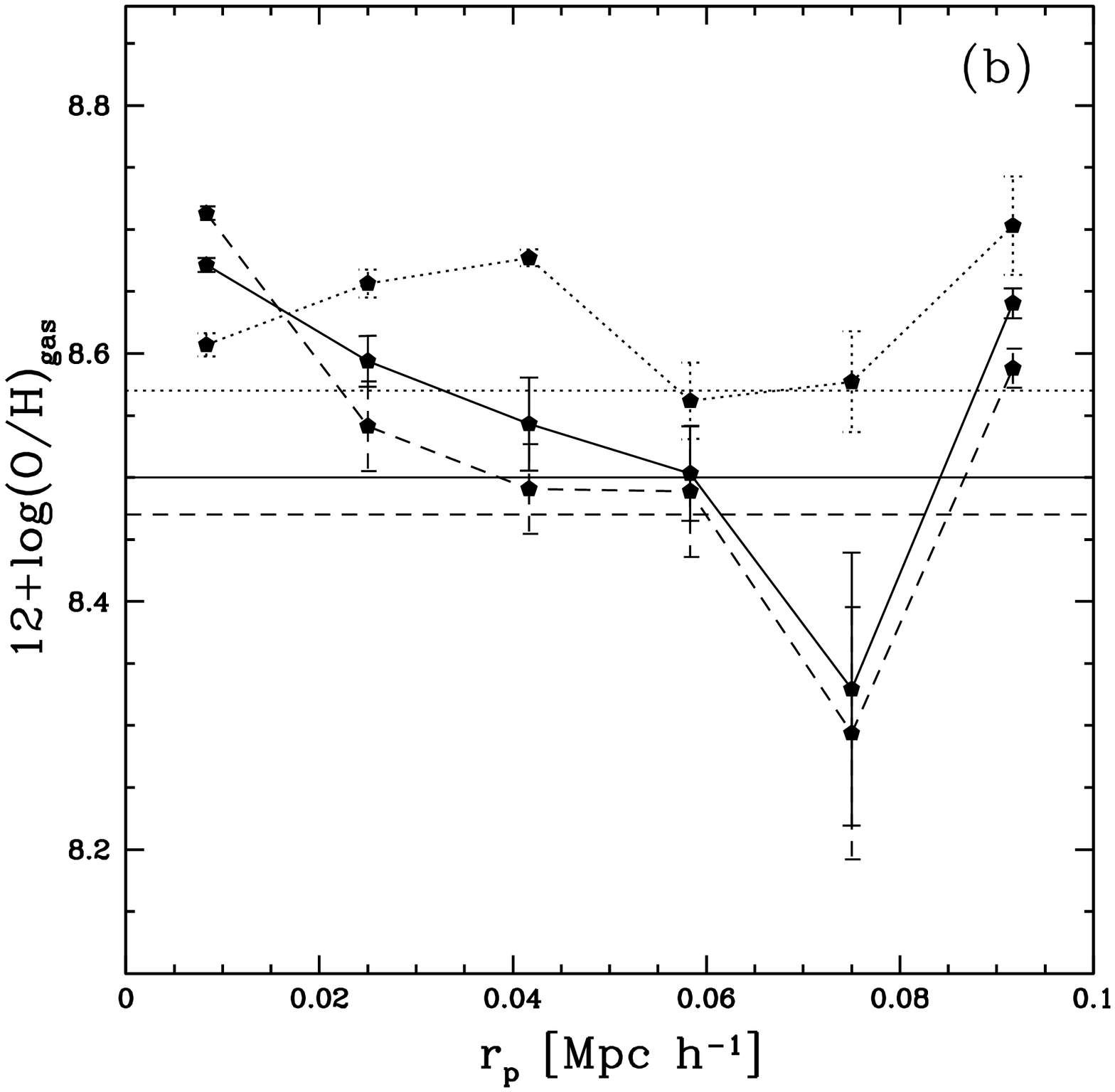}
\caption{Oxygen abundances for the stellar populations (a) and the interstellar medium (b)
as a function of projected distance to the closest neighbour, separated
in active SF (dotted lines), passive SF (dashed lines) and all (solid lines) galaxies  in pairs.
The corresponding mean abundances for galaxies in the control sample are shown as horizontal lines.
Error bars have been estimated by applying the boostrap resampling technique.}
\label{oxigen}
\end{figure}

In Fig.~\ref{oxigen}, we show the mean O/H 
 abundances of both the SPs (a) and the ISM (b)
of galaxies in pairs (solid lines) in the simulated 2D-GP catalog as a function of the 
projected separation.
Galaxies were also divided according to their star formation 
activity in  active star-forming (dotted lines) 
and passive star-forming  (dashed lines) systems.
The horizontal lines represent the 
corresponding  mean values for galaxies without a 
close companion divided in a similar fashion.

From Fig.~\ref{oxigen}a we  observe that the mean abundances of the SPs  in passive SF galaxies in pairs
  show an
enhancement of their chemical abundances with respect to the corresponding mean of the control sample for all projected distances.
Active SF galaxies in pairs show
an enhancement of the abundances of their SPs with respect to the mean abundance of the
control sample 
but at a lower level in comparison to passive SF systems.

\begin{figure}
\centering
\includegraphics[width=8cm,height=6.5cm]{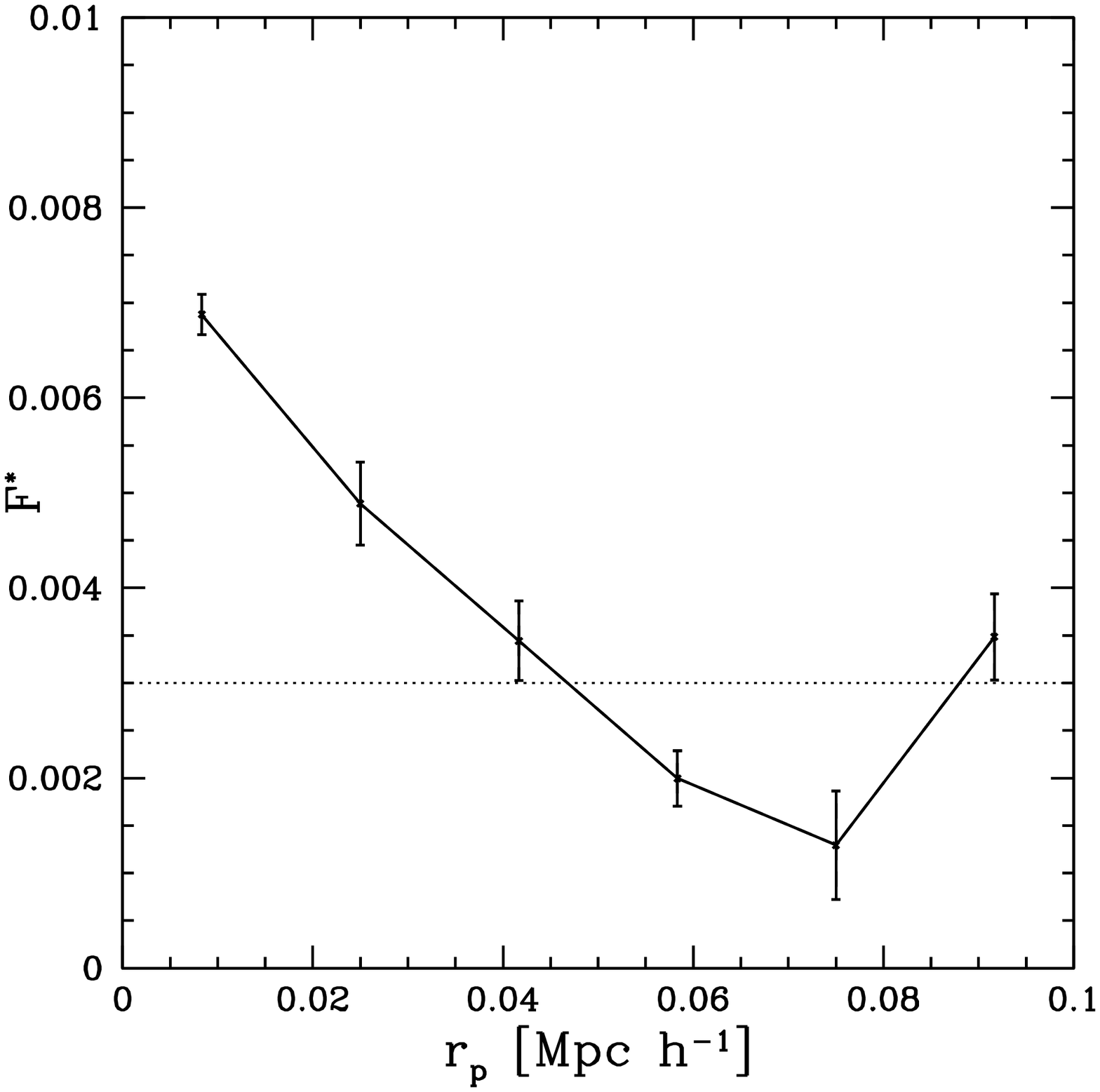}
\caption{Fraction ($F_{*}$) of stars formed in the last 0.5 Gyr as a function of the
relative separation for currently passive star forming galaxies in pairs. The horizontal
line represents the mean value of $F_{*}$ for the currently passive star forming galaxies without
a close companion. Boostrap errors are given.}
\label{fs-passive}
\end{figure}

In Fig.~\ref{oxigen}b  we can see  the abundances of the ISM for  galaxies in pairs. 
We find that, regardless of the presence of a close neighbour, active SF systems in pairs show  ISMs with higher
levels of enrichment.
On the other hand, currently passive star forming galaxies in pairs  show a clear 
correlation of their ISM  O/H abundances  with relative distance, 
fossils of recent  past interactions.
This correlation can be understood analysing the recent past SF history of
these systems. As it can be seen from Fig.~\ref{fs-passive},
the $F^*$  fraction for passive SF galaxies in pairs
anticorrelates with the relative distance to the nearest neighbour. 
Systems with a very close neighbour that have their gas components chemically enriched (Fig. ~\ref{oxigen}b),
 have experienced star formation within  the last 0.5 Gyr.

\begin{figure}
\centering
\includegraphics[width=8cm,height=6.5cm]{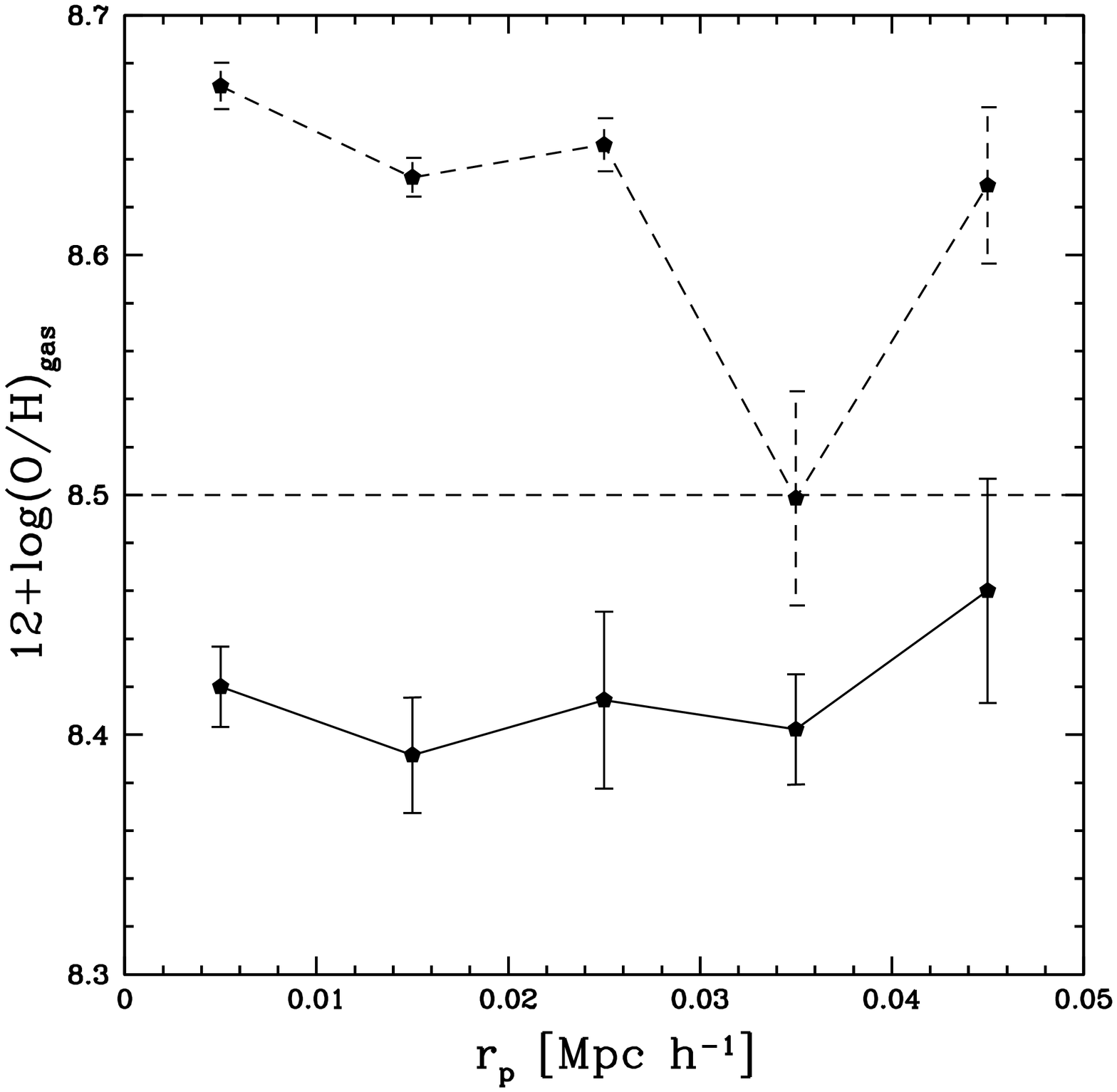}
\caption{Mean O/H abundances of the insterstellar medium in galaxies in pairs 
 estimated within 0.5 $r_{\rm opt}$ (solid line) and
2 $r_{\rm opt}$ (dashed line)  as a function of relative separation.
The horizontal line represents the mean O/H abundance for the control sample. }
\label{Ox_mstar}
\end{figure}

It is important to note that so far  O/H abundances have been estimated within two optical
radii providing information on the overall metallicity of baryons, in form of stars or gas, associated
to the simulated galaxies. Recent observational results on the metallicity properties of pairs by 
Kewley et al. (2005) showed a decrease
in the O/H abundance measured in the central regions 
of close galaxy pairs compared to galaxies without a close companion. These authors interpreted
this finding as a result of low metallicity gas inflows driven by the tidal torques generating during the interactions (e.g. Barnes \& Hernquist 1992, 1996).
In order to investigate this point in our simulations, we estimated the O/H abundances within the 
central region of the simulated galaxies defined within $0.5 \ r_{\rm opt}$ for close pairs.
In Fig. ~\ref{Ox_mstar} we have plotted the gas O/H abundances in the central regions (solid line) 
and within two optical radii (dashed line)
of galaxies in pairs with relative distances lesser than 50 kpc $h^{-1}$
 and the mean O/H abundance of galaxies in the control sample (horizontal line).
As it can be observed the central regions show lower level of enrichment with respect to
the mean O/H abundance of the control sample. 
The decrease in the mean O/H abundances
is produced by the contribution from the gas inflows from the external regions which also trigger
star formation activity as shown in Paper I.

\begin{figure}
\centering
\includegraphics[width=8cm,height=6.5cm]{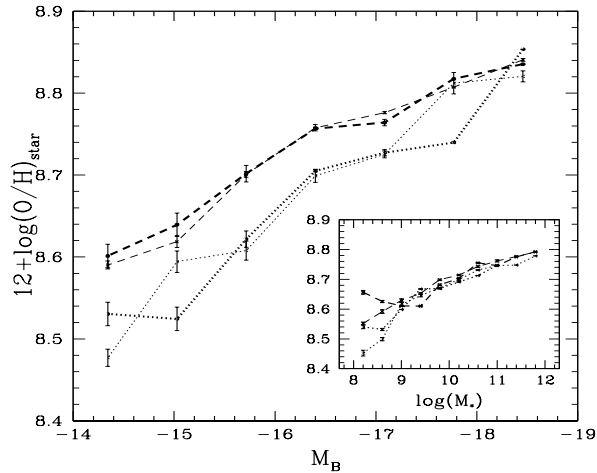}
\caption{Simulated luminosity-metallicity relation for active (dotted lines) and passive (dashed lines) SF systems in pairs (thick lines)
and in the control sample (thin lines). In the inset box we show the stellar mass-metallicity relation for
the same samples.}
\label{lmrnew}
\end{figure}

We also estimated the LMR for our simulated samples by using the overall mean stellar O/H abundances (i.e.
measured within $2 \ r_{\rm opt}$). Although galaxies with and without
a close companion exhibit the same trend for their LMR, a difference of 0.75 mag in the zero point of the LMRs is dectected if galaxies
are splitted according to their current SF activity into passive and active ones as shown by  
Fig. ~\ref{lmrnew}. We note that the simulated LMR has a similar slope to the observed relation 
although with a higher overall level of enrichment. A possible source for this
discrepancy could be due to the fact that we have used mean stellar O/H abundances for all simulated galaxies
 because they are less affected by 
numerical noise. Meanwhile, observations normally use abundances of the gas phase in HII regions of
 star forming galaxies. This does not invialidate our reasoning since both simulated samples have been selected from the
same simulation.
Nevertheless, the aim of this work is not to study the observed LMR but to assess if simulated galaxies
in pairs have a different LMR when compared to galaxies without a close companion.
The reader is refered to De Rossi et al. (in preparation) for an extensive discussion on 
 the LMR of galaxies in $\Lambda$-CDM scenarios.
Finally, we also estimated  the MMR for the same systems finding, in this case, no difference  for 
active or passive SF galaxies  (with and without a companion) 
as it can be appreciated in the inset box of Fig. ~\ref{lmrnew}\footnote{
The simulated MMR has a different slope than that reported by Tremonti et al. (2004) for
stellar masses lower than  $10^{10} M_{\odot} h^{-1}$.
 At lower masses, these simulations
show an excess of metals due to the lack of strong SN feedback (Tissera, De Rossi \& Scannapieco 2005).}.
Both results suggest that stellar mass is a more fundamental parameter than luminosity. 
In fact, this claim has been driven from   observational  (e.g. Tremonti et al. 2004; Erb et al. 2006)
and from numerical (Tissera et al. 2005) works. 
This lack of dependence of mean metallicity on stellar mass suggests that the displacement
of the LMR  for galaxies with different level of SF activity is 
 produced mainly  by the increase in the luminosity in those actively forming stars,  at a given mean stellar metallicity
level. 

\section{Conclusions}

In this work we analyse  galaxies in pairs in cosmological simulations in a $\Lambda$-CDM scenario as
a continuation of 
the work presented in  Paper I. 
We study colours and metallicity properties of  simulated galaxies with and without a close companion
in order to unveil the effects of galaxy-galaxy interactions.
We find that spurious pairs affect weakly the colours and the metallicity properties of galaxies in pairs.

We obtain that  interactions are able to induce a bimodal colour distribution  where the contribution
to   blue colours comes from those galaxies with recently significant star formation activity.
 Red galaxies
are formed, on average, by old stars ($\tau >$ 10 Gyr) which 
have experienced no significant star formation activity in the last 0.5 Gyr.
Galaxies without a close companion contribute with less stellar mass to the blue colours. 
This red excess is  produced by the  efficient transformation of 
gas into stars in our simulations due to the  lack of 
a self-consistent supernova feedback model that could help to regulate the star formation process (Scannapieco et al.
2006).
The analysis of merging and interacting pairs shows that the former contributes with a larger fraction of
blue stellar mass  than the latter, demostrating the ability of interactions in driving a colour bimodality.

The fraction of stellar mass in red galaxies 
increases with local density and this effect is  stronger if the galaxy has a close companion.
These results are consistent with our previous analysis of the SF activity in different environments 
discussed in Paper I which showed that the fraction of passive SF systems in pairs in high
density regions exceeds that observed in galaxies without a close companion.
 The excess of stellar mass with red colours in galaxies in  pairs detected
in our simulations is also
in agreement with recent observational results (De Propis et al. 2005; Alonso et al. 2006).

The ISMs of galaxies with active SF is highly enriched regardless of the presence of a close companion.
However, the abundances of the ISMs of galaxies in pairs with passive SF activity shows a dependence on  proximity to the nearest
neighbour. A similar dependence is found for the fraction of recently formed stars in these systems.
Hence, while the  chemical properties of galaxies with strong SF activity show an excess of metals
at all distances, 
those of passive SF systems anticorrelates with spatial proximity. This anticorrelation could 
store fossil records of  interactions and  could be, then,  used to unveil recent
SF episodes.
If central mean O/H abundances are measured, then, we detect a decrease in the metallicity level
for closer galaxy pairs produced by the increase of low metallicity gas inflows driven during the 
interactions. These gas inflows are  responsible of the increase in the star formation activity
reported in Paper I and studied in detail by  other authors (e.g. Mihos \& Hernquist 1996; Barnes \& Hernquist 1992, 1996; Tissera 2001;
Tissera et al. 2002).

If we use the abundances of the SPs as a chemical indicator, we found that the MMR of galaxies in pairs and galaxies without a close companion are very similar, regardless
of the SF activity of the systems. 
This finding suggests that the stellar mass is determining the metallicity content of a galaxy, at least,
when no strong galactic winds are present.
The LMR of galaxies in active SF in pairs and in the control sample
 show a displacement  towards brighter mangitudes by $\approx 0.75$ mag with respect to those
systems passively forming stars.
The results obtained by using the chemical abundances of the ISM as a chemical indicator are
quite noise making difficut to draw any conclusion.

Our analysis suggests that galaxy-galaxy interactions can significantly contribute to the 
 determination of the bimodal colour galaxy distribution and chemical relations.
In particular, the strong dependence of the astrophysical properties of galaxies in pairs on environment
suggests that interactions are an efficient process of transforming galaxies considering that 
the rates of mergers and interactions increase with local density  in 
hierarchical clustering scenarios.
 The agreement between these results and recent observational findings
 supports the current cosmological paradigm for galaxy formation
while the discrepancies suggest the need for strong SN feedback.

\begin{acknowledgements}
P.B.Tissera thanks S. Charlot for making publically available specific subroutines that helped to 
couple the GALAXEV with the results from our simulations. 
 C. Scannapieco thanks the Alexander von Humboldt
Foundation, the Federal Ministry of Education and Research and the
Programme for Investment in the Future (ZIP) of the German Government
for partial support. P.B.Tissera thanks the Aspen Center for Physics
during the 2004 Summer Workshop for useful discussions that help to the development of this work.
The simulations were run on  Ingeld PC-cluster funded by Fundaci\'on Antorchas.
This work was partially supported by the
Consejo Nacional de Investigaciones Cient\'{\i}ficas y 
T\'ecnicas, Fundaci\'on Antorchas  (Argentina) 
and by the European Union's ALFA-II
programme, through LENAC, the Latin American European Network for
Astrophysics and Cosmology.
\end{acknowledgements}

\end{document}